\documentclass[10pt,letterpaper]{article}
\usepackage{subfig}
\usepackage{amsmath,amssymb}
\usepackage{cite}
\usepackage{epstopdf}

\usepackage{opex3}

\begin{document}

\title{Experimental compressive phase space tomography}

\author{Lei Tian,$^{1,*}$ Justin Lee,$^{2}$ Se Baek Oh,$^{1,4}$ and George Barbastathis$^{1,3}$}

\address
{$^1$Department of Mechanical Engineering, Massachusetts Institute of Technology, \\ 77 Massachusetts Avenue, Cambridge, MA 02139, USA \\
$^2$Department of Health Science and Technology, Massachusetts Institute of Technology, \\77 Massachusetts Avenue, Cambridge, MA 02139, USA \\
$^3$Singapore-MIT Alliance for Research and Technology (SMART) Centre, \\3 Science Drive 2, Singapore 117543, Singapore\\
$^4$Currently with KLA-Tencor, 1 Technology Drive, Milpitas, CA 95035, USA \\}

\email{$^*$lei\_tian@mit.edu}

\begin{abstract}
Phase space tomography estimates correlation functions entirely from snapshots in the evolution of the wave function along a time or space variable. In contrast, traditional interferometric methods require measurement of multiple two--point correlations. However, as in every tomographic formulation, undersampling poses a severe limitation. Here we present the first, to our knowledge, experimental demonstration of compressive reconstruction of the classical optical correlation function, {\it i.e.} the mutual intensity function. Our compressive algorithm makes explicit use of the physically justifiable assumption of a low--entropy source (or state.) Since the source was directly accessible in our classical experiment, we were able to compare the compressive estimate of the mutual intensity to an independent ground--truth estimate from the van Cittert--Zernike theorem and verify substantial quantitative improvements in the reconstruction. 
\end{abstract}



\section{Introduction}
Correlation functions provide complete characterization of wave fields in several branches of physics, e.g. the mutual intensity of stationary quasi--monochromatic partially coherent light~\cite{mandel1995optical}, and the density matrix of conservative quantum systems ({\it i.e.}, those with a time--independent Hamiltonian)~\cite{densitymatrix}. Classical mutual intensity expresses the joint statistics between two points on a wavefront, and it is traditionally measured using interferometry: two sheared versions of a field are overlapped in a Young, Mach--Zehnder, or rotational shear~\cite{Itoh:86, Marks:99} arrangement, and two--point ensemble statistics are estimated as time averages by a slow detector under the assumption of ergodicity~\cite{mandel1995optical, 2000stop.book.....G}. 

As an alternative to interferometry, phase space tomography (PST) is an elegant method to measure correlation functions. In classical optics, PST involves measuring the intensity under spatial propagation~\cite{PhysRevLett.68.2261, PhysRevLett.72.1137, Tran:05} or time evolution~\cite{Beck:93}. In quantum mechanics, analogous techniques apply~\cite{1989PhRvA..40.2847V, PhysRevLett.70.1244, PhysRevLett.74.4101, 1997Natur.386..150K}. However, the large dimensionality of the unknown state makes tomography difficult. In order to recover the correlation matrix corresponding to just $n$ points in space, a standard implementation would require at least $n^2$ data points. 

Compressed sensing~\cite{candes2006robust, Candes2006, Donoho:2006sf} exploits sparsity priors to recover missing data with high confidence from a few measurements derived from a linear operator. Here, sparsity means that the unknown vector  contains only a small number of nonzero entries in some specified basis. Low--rank matrix recovery~(LRMR)~\cite{Candes:fk, Candes:2010:PCR:1823677.1823678} is a generalization of compressed sensing from vectors to matrices: one attempts to  reconstruct a high--fidelity and low--rank description of the unknown matrix from very few linear measurements. 

In this paper, we present the first, to our knowledge, experimental measurement and verification of the correlation function of a classical partially coherent field using LRMR. It is worth noting that LRMR came about in the context of compressive quantum state tomography (QST)~\cite{PhysRevLett.105.150401}, which utilizes different physics to attain the same end goal of reconstructing the quantum state. In PST, one performs tomographic projection measurements, rotating the Wigner space between successive projections by evolving the wave function~\cite{PhysRevLett.68.2261, PhysRevLett.72.1137}. This is directly analogous to the classical optical experiment we are presenting here, where we perform intensity measurements ({\it i.e.}, tomographic projections in Wigner space) and utilize propagation along the optical axis to rotate the Wigner space between projections. The difference lies in the fact that in QST the state is recovered via successive applications of the Pauli dimensionality--reducing operator, and there is no need to evolve the state. Nevertheless, both approaches lead to the same Hermitian LRMR problem, as long as the assumption of a quasi--pure unknown state is satisfied. In~\cite{5714248}, it was shown that estimation of a low--rank matrix of dimension $n$ and rank $r$ requires only $O(rn\ln n)$ to $O(rn\ln^2 n)$ data points.  A similar LRMR method was also used to recover the complex amplitude of an unknown object under known illumination~\cite{Candes:2011fk, 2011arXiv1109.0573C,Shechtman:11}. Since the complex amplitude of the object is time--invariant, a rank--one solution was assumed in these works.

The low--rank assumption for classical partially coherent light anticipates a source composed of a small number of mutually incoherent effective sources, {\it i.e.} coherent modes~\cite{Wolf:82}, to describe measurements. This is essentially equivalent to the low entropy assumption~\cite{PhysRevLett.105.150401}, {\it i.e.} a nearly pure quantum state in the quantum analogue. This assumption is valid for lasers, synchrotron and table--top X--ray sources~\cite{Pelliccia:11}, and K\"{o}hler illumination in optical microscopes~\cite{mandel1995optical}. An additional requirement for LRMR to succeed is that measurements are ``incoherent'' with respect to the eigenvectors of the matrix, {\it i.e.} the measured energy is approximately evenly spread between modes~\cite{5714248, Candes:2011fk}. Diffraction certainly mixes the coherent modes of the source rapidly, so we expect LRMR to perform well for classical PST. The same expectation for QST has already been established~\cite{PhysRevLett.105.150401}.

\section{Theory and method}
The two--point correlation function of a stationary quasi--monochromatic partially spatially coherent field is the mutual intensity function~\cite{mandel1995optical}
\begin{equation}
J(x_1, x_2) = \left<g^*(x_1)g(x_2)\right>,
\label{eq:MI}
\end{equation}
where $\left<\cdot\right>$ denotes the expectation value over a statistical ensemble of realizations of the field $g(x)$.

The measurable quantity of the classical field, {\it i.e.} the intensity, after propagation by distance $z$, is~\cite{mandel1995optical}
\begin{equation}
I(x_{\text{o}};z) = \iint dx_1dx_2 J(x_1, x_2) 
\exp\left(-\frac{i\pi}{\lambda z}(x_1^2-x_2^2)\right)\exp\left(i2\pi\frac{x_1-x_2}{\lambda z}x_{\text{o}}\right).
\label{eq:J_tr} 
\end{equation}
This can be expressed in operator form as
\begin{equation}
I = \text{tr}(P_{x_{\text{o}}}J), 
\end{equation}
where $P$ denotes the free--space propagation operator that combines both the quadratic phase and Fourier transform operations in Eq.~(\ref{eq:J_tr}), tr$(\cdot)$ computes the trace, and $x_{\text{o}}$ denotes the lateral coordinate at the observation plane. By changing variables $x = \left(x_1+x_2\right)/2$, $x' = x_1-x_2$ and Fourier transforming the mutual intensity with respect to $x$ we obtain the Ambiguity Function (AF)~\cite{Brenner1983323, 1984AcOpt..31..213B, Tu:1997fk} 
\begin{eqnarray}
\mathcal{A}\left(u',x'\right) =\int J\left( x+\frac{x'}{2}, x-\frac{x'}{2} \right) \exp\left(-i2\pi u'x \right)dx.
\label{eq:AF}
\end{eqnarray}  
Eq.~(\ref{eq:J_tr}) can be written as~\cite{PhysRevLett.68.2261, PhysRevLett.72.1137, Tran:05, Brenner1983323, Tu:1997fk},
\begin{equation}
\tilde{I}(u';z) = \mathcal{A}\left(u',\lambda zu'\right),
\label{eq:afslice}
\end{equation}
where $\tilde{I}$ is the Fourier transform of the vector of measured intensities with respect to $x_{\text{o}}$. Thus, radial slices of the AF may be obtained from Fourier transforming the vectors of intensities measured at corresponding propagation distances, and from the AF the mutual intensity can be recovered by an additional inverse Fourier transform, subject to sufficient sampling.

To formulate a linear model for compressive PST, the measured intensity data is first arranged in Ambiguity space. The mutual intensity is defined as the ``sparse'' unknown to solve for. To relate the unknowns (mutual intensity) to the measurements (AF), the center--difference coordinate--transform is first applied, expressed as a linear transformation $T$ upon the mutual intensity $J$, followed by Fourier transform $\mathcal{F}$, and adding measurement noise $e$ as
\begin{equation}
\mathcal{A}= \mathcal{F}\cdot T\cdot J +e.
\end{equation}

The mutual intensity propagation operator is unitary and Hermitian, since it preserves energy. We use eigenvalue decomposition to determine the basis where the measurement is sparse. The resulting basis, {\em i.e.} the set of eigenvectors, is also known as coherent modes in optical coherence theory, whereas the whole process is known as coherent mode decomposition~\cite{Wolf:82}. The goal of the LRMR method is to minimize the number of coherent modes to describe measurements. By doing LRMR,  we impose two {\em physically} meaningful priors: {\em (1)} existence of the coherent modes~\cite{Wolf:82}, and {\em (2)} sparse representation of the partially coherent field in terms of coherent modes. 

Mathematically, if we define all the eigenvalues $\lambda_i$ and the estimated mutual intensity as $\hat{J}$, the method can be written as 
\begin{eqnarray}
& \text{minimize  } &  ~\text{rank}(\hat{J}) \nonumber\\
& \text{subject to  }  &~\mathcal{A}=\mathcal{F}\cdot T\cdot \hat{J}, \nonumber \\
& & \lambda_i\geq0, \text{and } \sum_i \lambda_i = 1.
\end{eqnarray} 
Direct rank minimization is NP--hard; however, it can be accomplished by solving instead a proxy problem: convex minimization of the ``nuclear norm'' ($\ell_1$ norm) of the matrix $J$~\cite{Candes:fk, 2009arXiv0903.3131C}. 
The corresponding problem is stated as 
\begin{eqnarray}
& \text{minimize  } &  ~\parallel\hat{J}\parallel_* \nonumber\\
& \text{subject to  }  &~\mathcal{A}=\mathcal{F}\cdot T\cdot \hat{J}, \nonumber \\
& & \lambda_i\geq0, \text{and } \sum_i \lambda_i = 1,
\end{eqnarray} 
where the nuclear norm is the sum of the singular values $\sigma_i=|\lambda_i|$, $\parallel\hat{J}\parallel_*=\sum_i\sigma_i$. This problem is convex and a number of numerical solvers can be applied to solve it. In our implementation, we used the singular value thresholding (SVT) method~\cite{2008arXiv0810.3286C}. The output estimate after each iteration of SVT typically has a sub-normalized total energy, {\it i.e.} $\sum_i\lambda_i<1$; we compensated for this by renormalizing at the end of each iteration~\cite{PhysRevLett.105.150401}.

\section{Numerical simulations }

\begin{figure}[htb]
\centering
\subfloat[]{\includegraphics[width=0.35\textwidth]{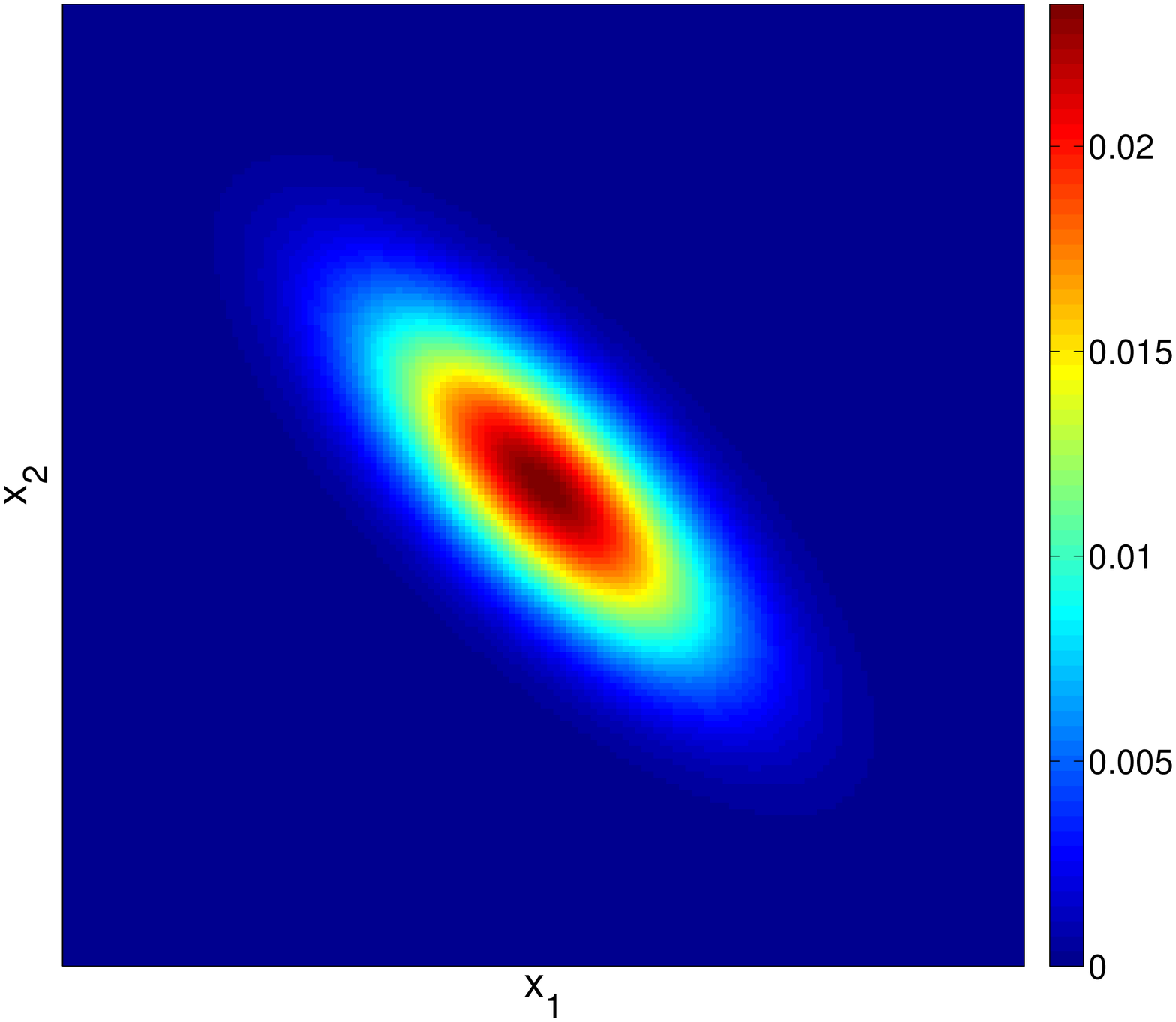}}
\subfloat[]{\includegraphics[width=0.35\textwidth]{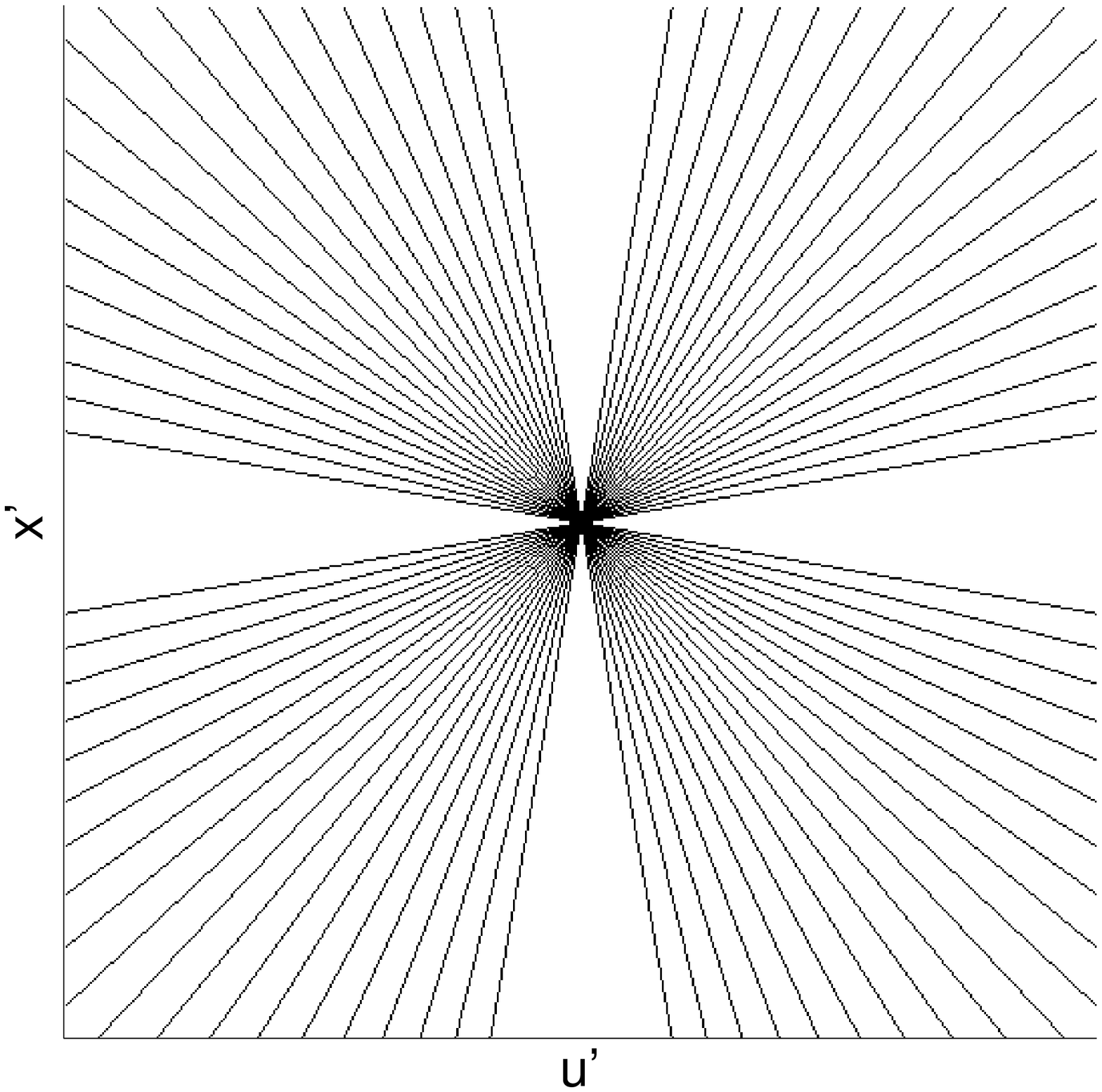}}

\subfloat[]{\includegraphics[width=0.35\textwidth]{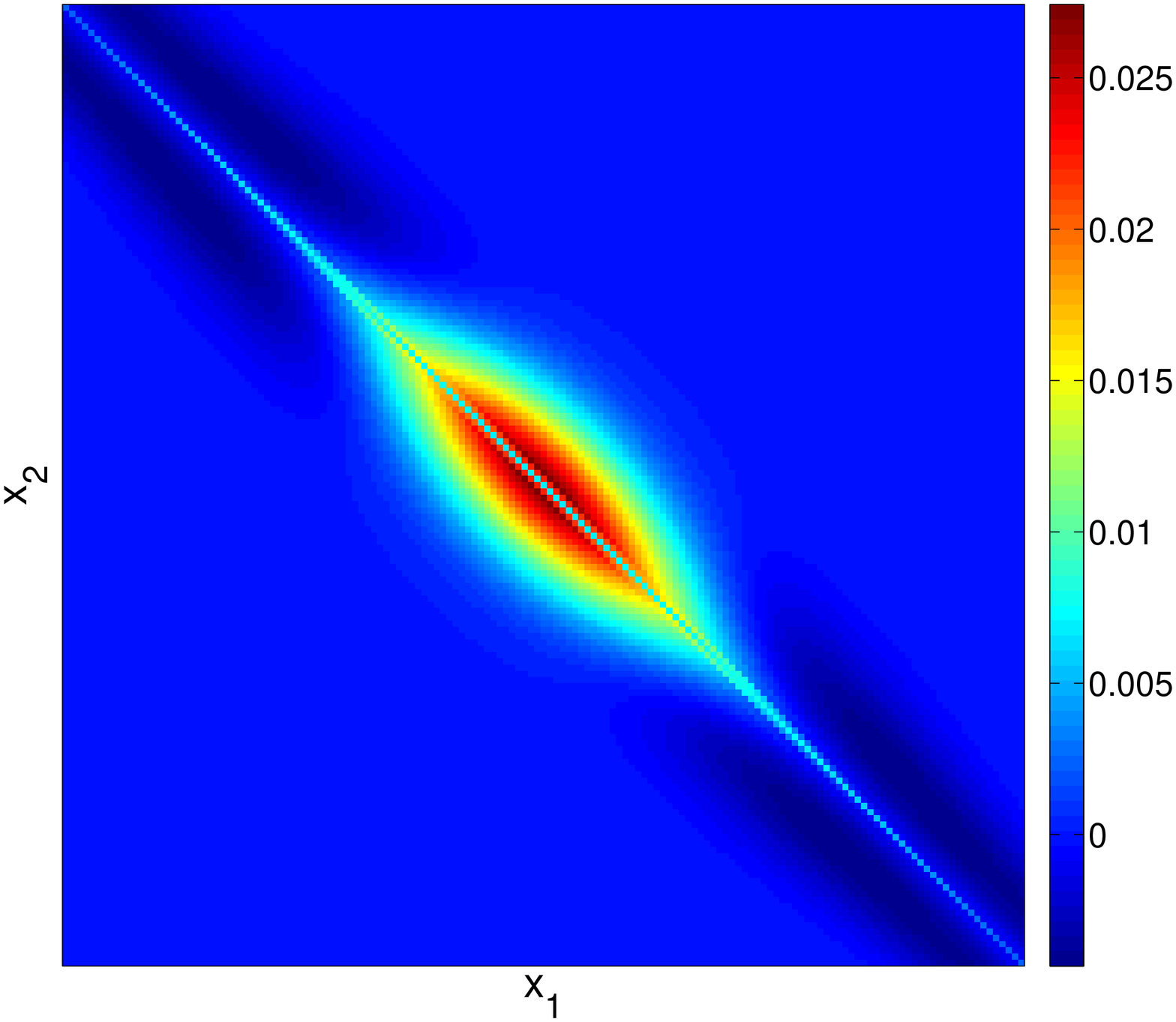}}
\subfloat[]{\includegraphics[width=0.35\textwidth]{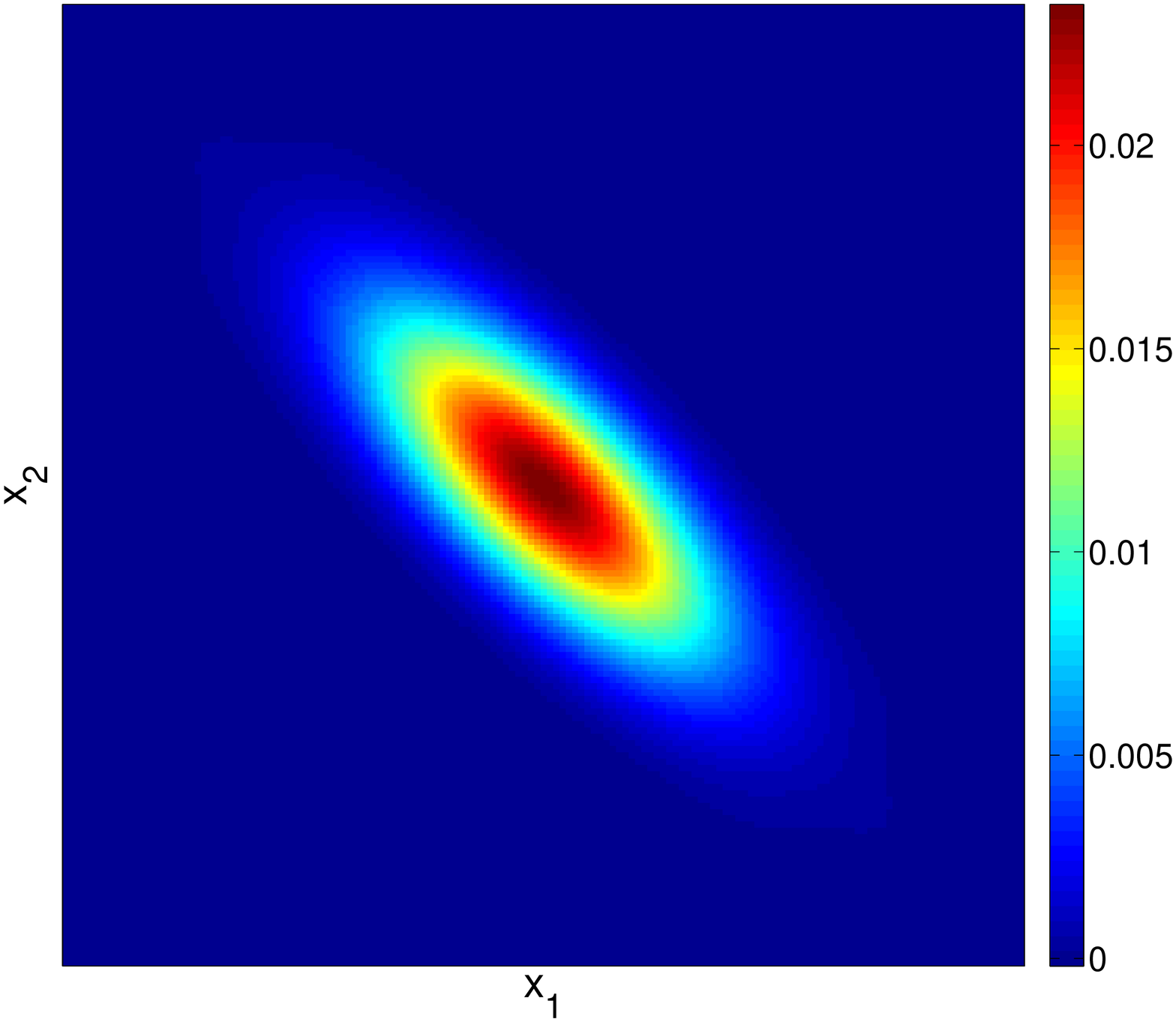}}
\caption{(a) Input mutual intensity of a GSMS with paramters $\sigma_I = 17$ and $\sigma_c = 13$, (b) data point locations in the Ambiguity space, mutual intensities estimated by (c) FBP and (d) LRMR methods.}
\label{fig:MI_simulation}
\end{figure}

\begin{figure}[htb]
\centering
\subfloat[]{\includegraphics[width=0.5\textwidth]{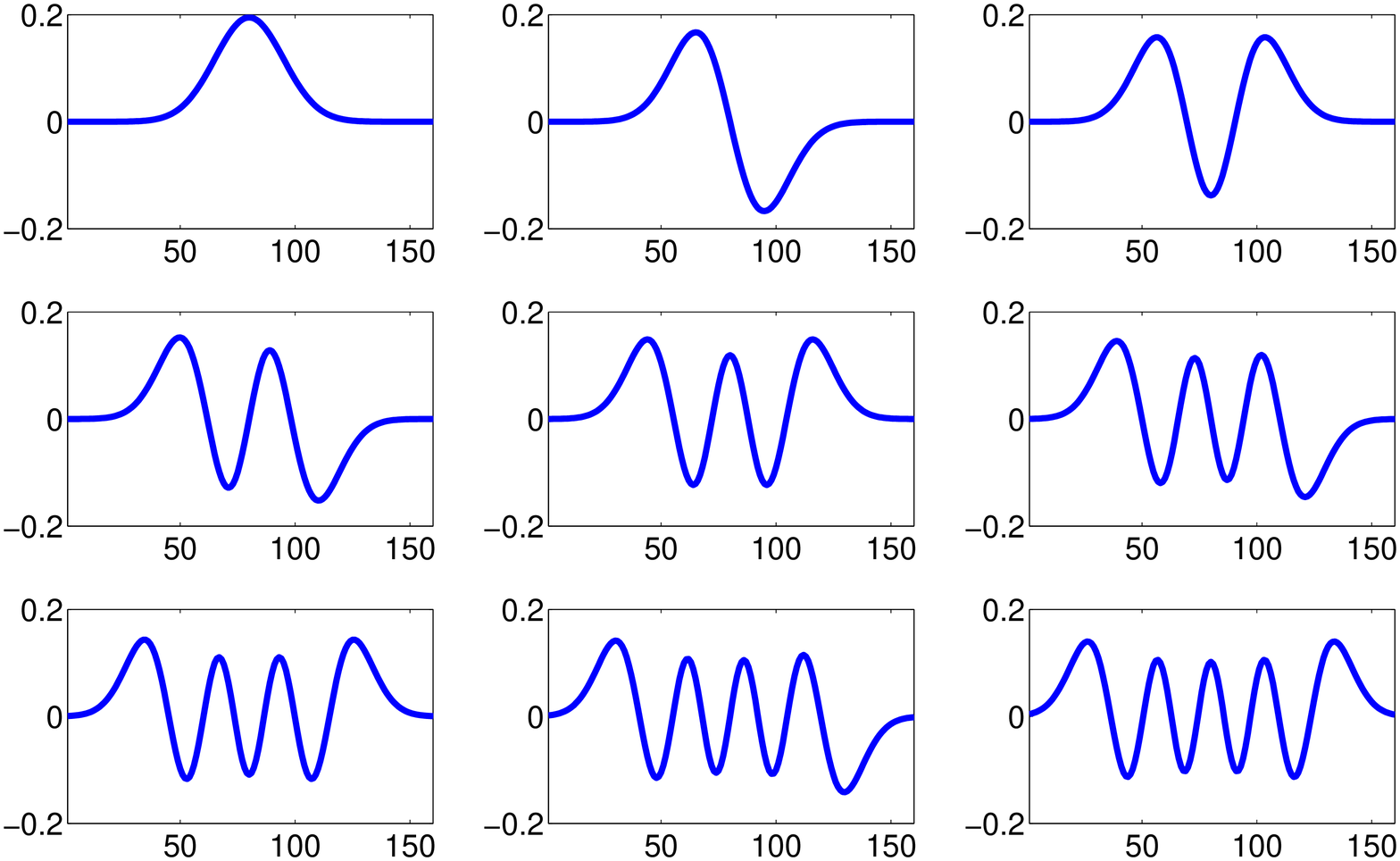}}
\subfloat[]{\includegraphics[width=0.5\textwidth]{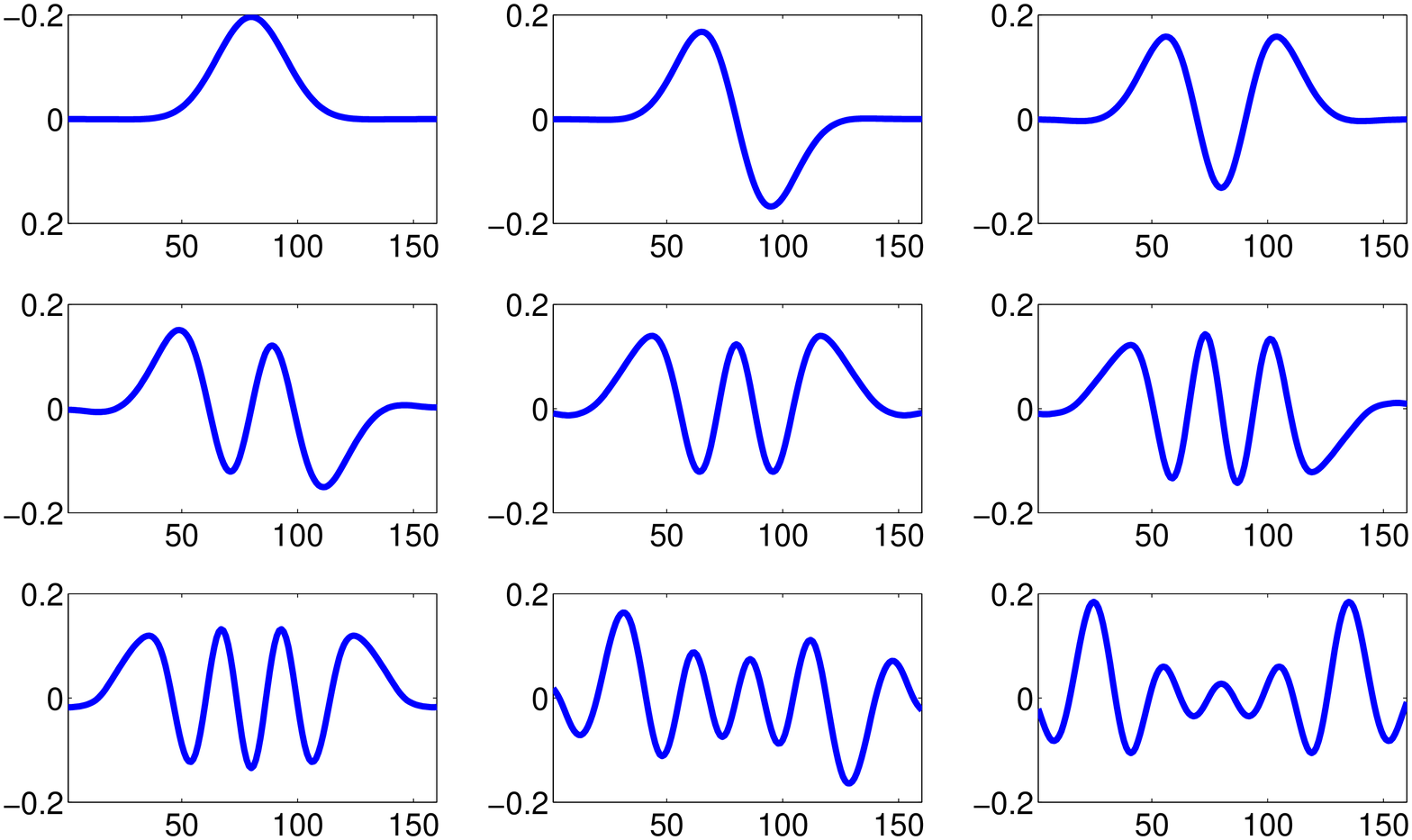}}
\caption{The first nine coherent modes of the mutual intensity in Fig.~\ref{fig:MI_simulation}(a). (a) Theoretical modes, and (b) LRMR estimates.}
\label{fig:Mode_simulation_vec}
\end{figure}

\begin{figure}[!]
\centering
\subfloat[]{\includegraphics[width=0.4\textwidth]{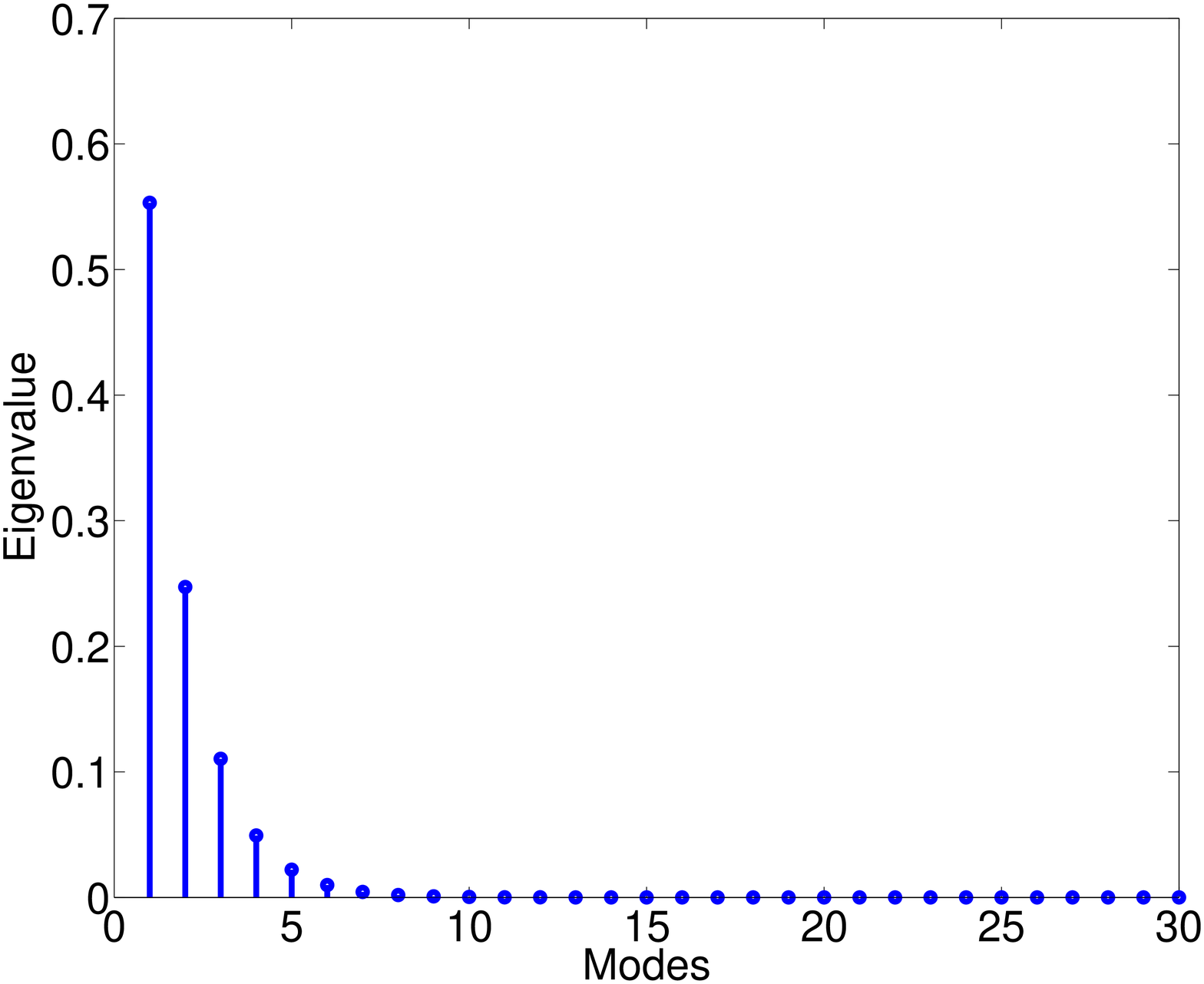}}
\subfloat[]{\includegraphics[width=0.4\textwidth]{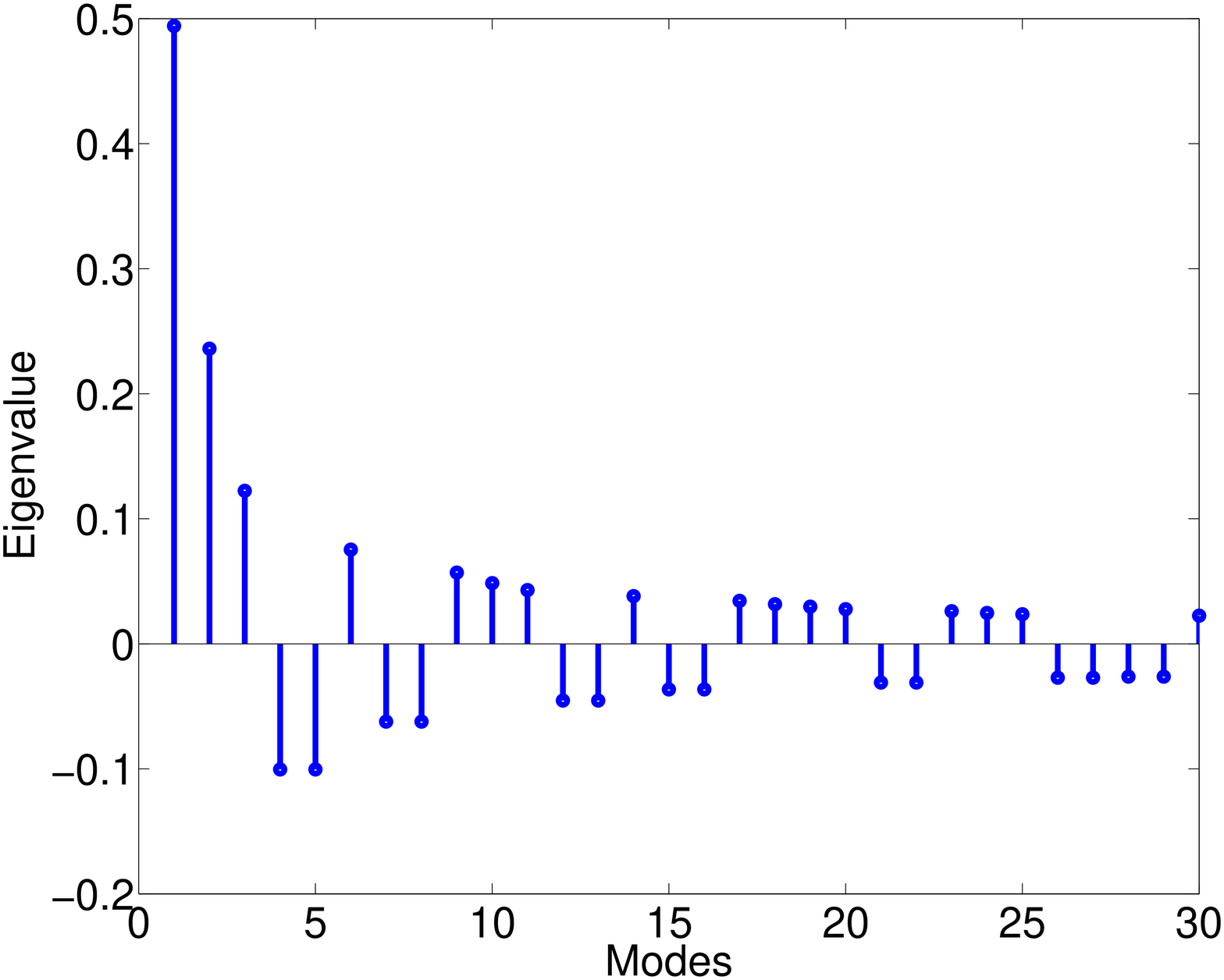}}

\subfloat[]{\includegraphics[width=0.4\textwidth]{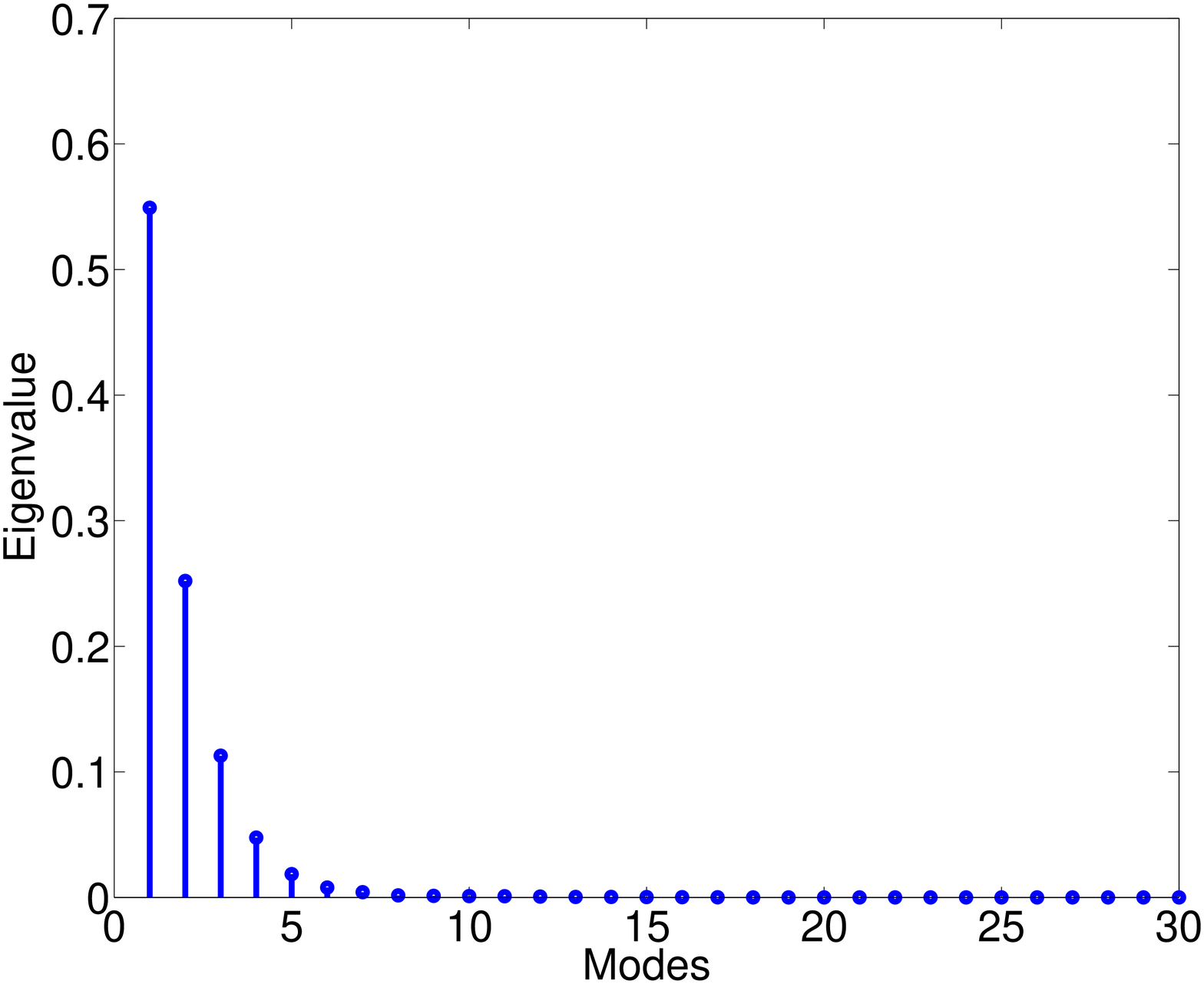}}
\subfloat[]{\includegraphics[width=0.4\textwidth]{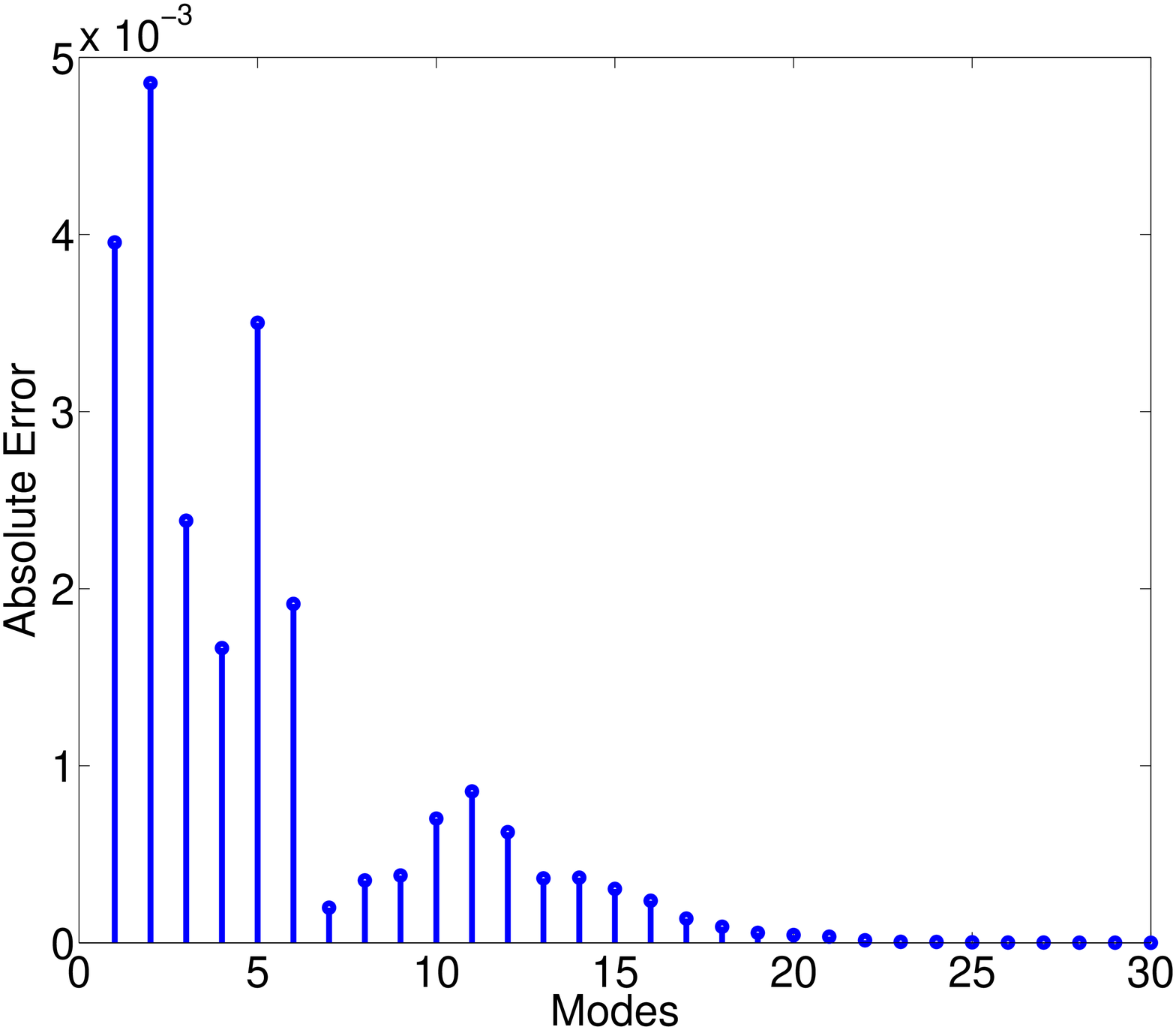}}
\caption{ Eigenvalues of the mutual intensity in Fig.~\ref{fig:MI_simulation}(a). (a) Theoretical values, (b) FBP estimates,  (c) LRMR estimates, and (d) absolute errors in the LRMR estimates versus mode index. }
\label{fig:Mode_simulation_val}
\end{figure}

First, we demonstrate the LRMR method with a numerical example using a 1D Gaussian-Schell model source~(GSMS). Both the intensity distribution and the degree of coherence of GSMS follow a Gaussian distribution~\cite{Starikov:82}
\begin{equation}
J\left(x_1,x_2\right) = [I(x_1)]^{1/2}[I(x_2)]^{1/2}\mu(x_1-x_2), 
\end{equation}
and
\begin{eqnarray}
& & I(x) = \exp\left(\frac{x^2}{2\sigma_I^2}\right), \nonumber \\
& & \mu(x_1-x_2) = \exp\left(\frac{(x_1-x_2)^2}{2\sigma_c^2}\right),
\end{eqnarray}
where $\sigma_I$ determines the spatial extent of the source, and $\sigma_c$ is proportional to the coherence length and determines the number of coherent modes in the input source. The eigenvalues of GSMS are never zero (analytical solution given in~\cite{Starikov:82}). We defined the number of modes (rank of the source) $r$ as the first $r$ modes containing the $99\%$ of the total energy.

One example is shown in Fig.~\ref{fig:MI_simulation}(a). The parameters in this example are $\sigma_I = 17$ and $\sigma_c = 13$ (rank $r=6$). Intensities are calculated at 40 different axial distances and the coverage in Ambiguity space is shown in Fig.~\ref{fig:MI_simulation}(b). We simulate the case where data from both the near field and the far field are missing due to the finite range of camera scanning motion allowed in the actual experiment. The missing cone around the $u'$-axis is due to missing data from near field, while the data missing from far field results in the missing cone around the $x'$-axis. Both cones have an apex angle of 20 degrees. 

For comparison, the data is first processed using the traditional filtered-backprojection (FBP) method~\cite{Kak:1988fk}. Applying the Fourier-slice theorem to Eq.~(\ref{eq:afslice}) implies that the 1D Fourier transform of a radial slice in the Ambiguity space (an intensity measurement) is related to a projection in the AF's 2D Fourier space (the Wigner space~\cite{1978OptCo..25...26B,Bastiaans:86}). The Wigner distribution function (WDF) is related to the mutual intensity by
\begin{equation}
\mathcal{W}\left(x,u\right) =\int J\left( x+\frac{x'}{2}, x-\frac{x'}{2} \right) \exp\left(-i2\pi ux' \right)dx'.
\end{equation}
To implement the FBP method, each intensity projection is first filtered by a Ram--Lak kernel apodized by a Hamming window; the estimated WDF is obtained by back-projecting all the filtered intensities, and then an inverse Fourier transform is applied to produce the mutual intensity. Figure~\ref{fig:MI_simulation}(c) shows the reconstructed mutual intensity following this procedure. Three types of artifacts can be seen in this reconstruction.  First, the reconstructed mutual intensity has lower values along the diagonal of the matrix due to the missing cones. However, this is unphysical because a correlation function should always have maximum value at zero separation.  Second, the estimated degree of coherence is lower than the original field. The third artifact is the high frequency noise around the diagonal of the matrix, which is due to undersampling between the radial slices. All these artifacts have been greatly suppressed or completely removed by LRMR, whose reconstruction result is shown in Fig.~\ref{fig:MI_simulation}(d). The disappearance of the correlation peak along the diagonal ({\it i.e.}, the intensity) when we use FBP for the reconstruction can be best explained with the help of Figure~\ref{fig:MI_simulation}(b). Going from the Ambiguity space to the mutual intensity space involves Fourier transforming along horizontal lines, parallel to the $u'$ axis. The diagonal in particular corresponds to the line $x'=0$. It can be easily seen that, due to the missing cone, pretty much all the data are missing from that line, except near the origin; thus resulting in a low--pass filtering effect. The fact that the compressive reconstruction method manages to restore the physically correct values of the correlation along the diagonal corroborates that the missing cone is successfully retrieved in our LRMR reconstruction. The FBP reconstruction may also be compared quantitatively to the compressive reconstruction in terms of the global degree of coherence parameter $\bar{\mu} = \frac{\sqrt{\sum_i \lambda_i^2}}{\sum_i |\lambda_i|}$~\cite{Starikov:82mode, Bastiaans:84}, which was found as 0.150 and 0.617, respectively; the true state has $\bar{\mu} = 0.618$.

The coherent modes for a GSMS are Hermite--Gaussian sources~\cite{Starikov:82}. The theoretical and LRMR estimated first nine coherent modes in this example are shown in Fig.~\ref{fig:Mode_simulation_vec}(a) and \ref{fig:Mode_simulation_vec}(b), respectively. The theoretical eigenvalues are shown in Fig.~\ref{fig:Mode_simulation_val}(a). The FBP and LRMR estimated eigenvalues are compared in Fig.~\ref{fig:Mode_simulation_val}(b) and \ref{fig:Mode_simulation_val}(c), respectively. The FBP estimates have several negative values, which does not satisfy the positive energy constraint. The absolute errors in LRMR estimates are plotted in Fig.~\ref{fig:Mode_simulation_val}(d).

\begin{figure}[htb]
\centering
\subfloat[Gaussian noise]{\includegraphics[width=0.5\textwidth]{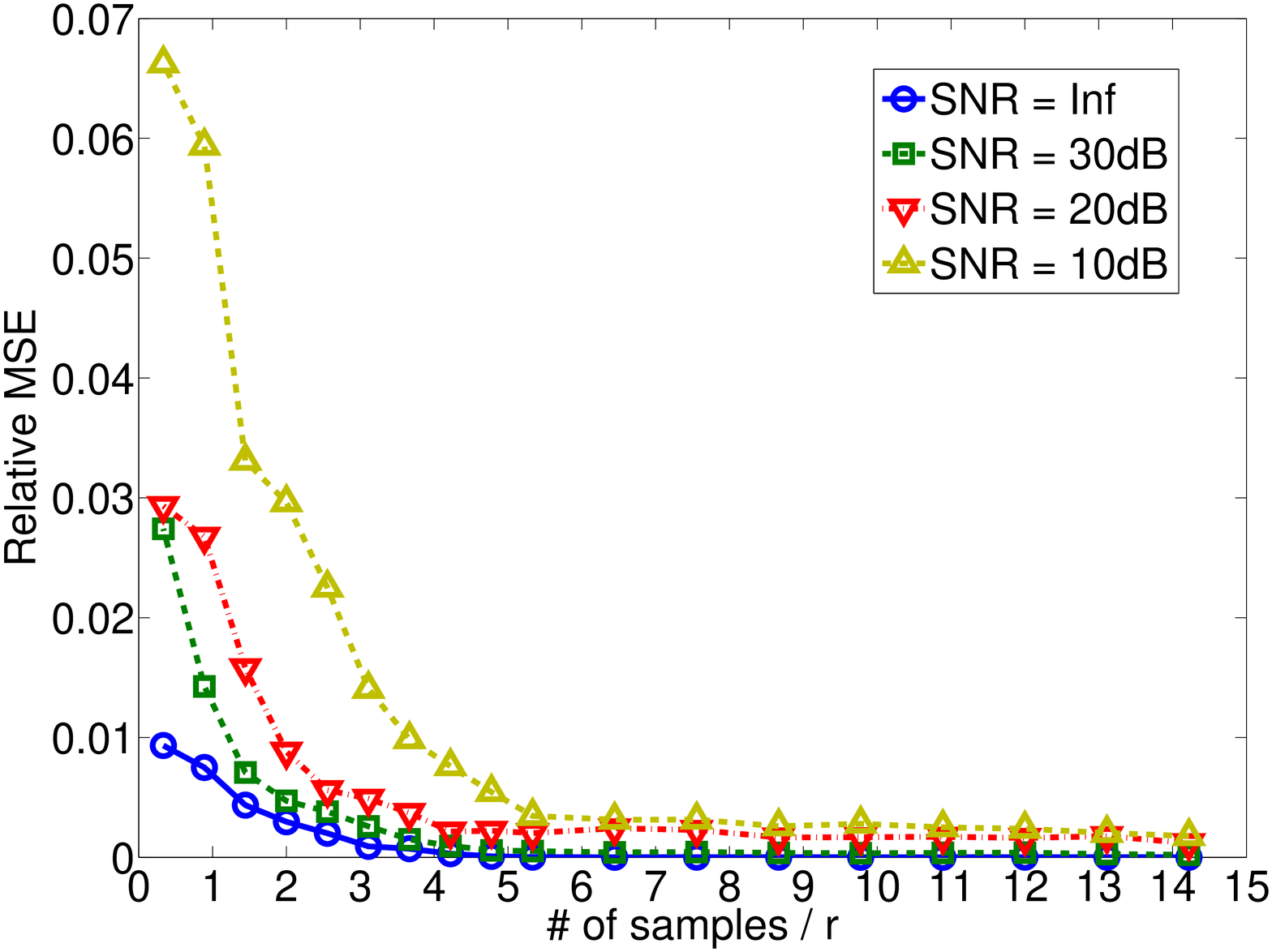}}
\subfloat[Poisson noise]{\includegraphics[width=0.5\textwidth]{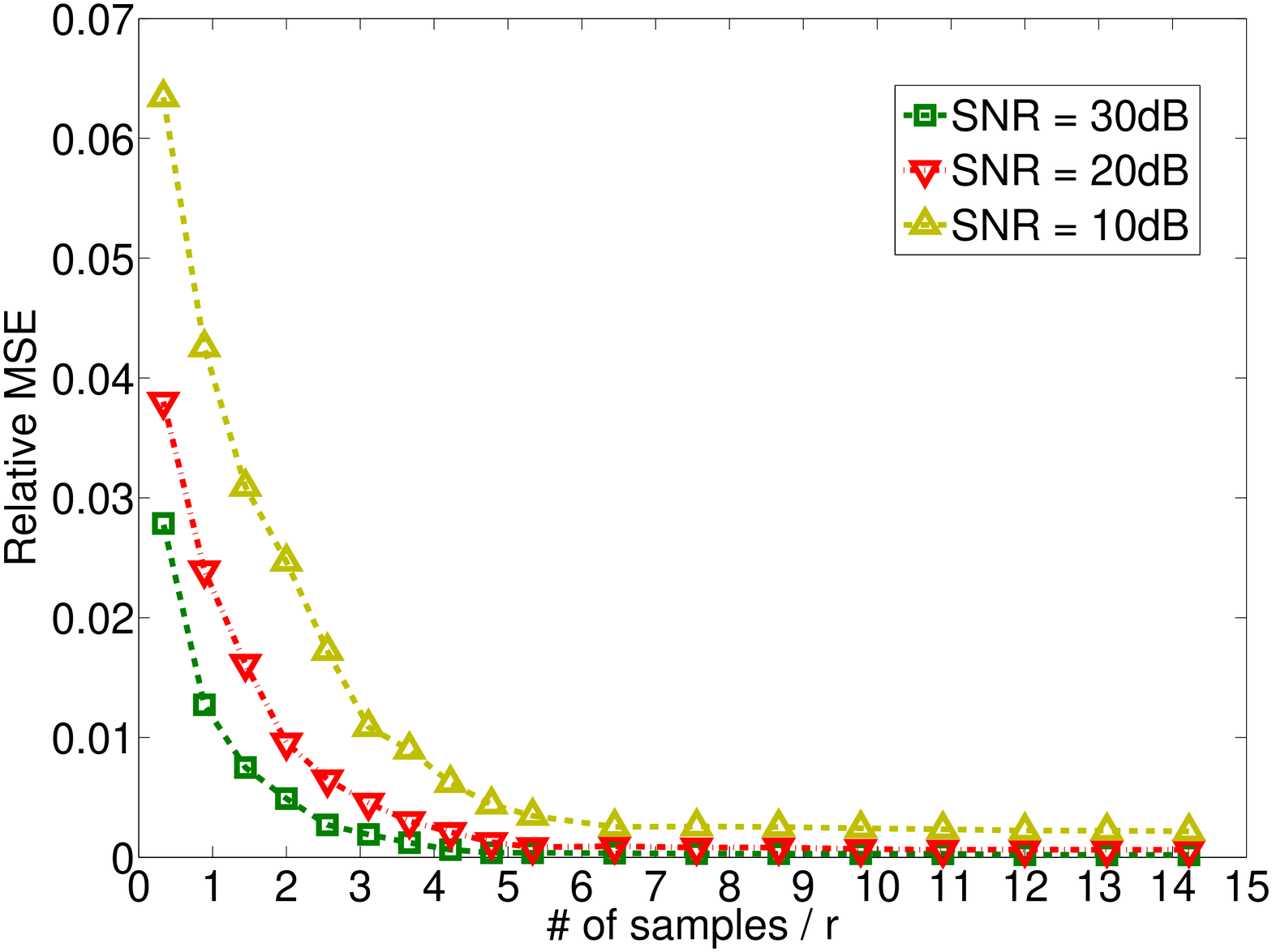}}
\caption{Oversampling rate versus relative MSE of LRMR estimates. The input field is a GSMS with parameters $\sigma_I = 36$ and $\sigma_c = 18$. The noisy data is generated with different SNR from (a) an additive random Gaussian noise model, and (b) a Poisson noise model. }
\label{fig:SNR}
\end{figure}

Next, we study the noise performance of the LRMR method with a numerical example. In this simulation, the dimension of the input GSMS is $256\times 256$ with parameters $\sigma_I = 36$ and $\sigma_c = 18$ (rank $r=9$). We generate noisy data with different signal-to-noise ratio~(SNR) from both an additive random Gaussian noise model and a Poisson noise model. However, we emphasize that the reconstruction algorithm does not make use of the noise statistics. For each SNR level, we repeat the simulation 100 times with different random noise terms, and then record the average relative mean-square-error (MSE) from the LRMR reconstruction. The ratio between the number of samples taken from the intensity measurements and the rank $r$ of the input mutual intensity matrix determines the oversampling rate~\cite{2011arXiv1109.0573C}. This rate is plotted versus relative MSE for different SNR cases in Fig.~\ref{fig:SNR}. For good performance, the required oversampling rate is at least 5--6 (the theoretical oversampling rate is on the order of $\ln(256)=5.5$ according to ~\cite{5714248}). Furthermore, the LRMR method is robust to noise in the sense that the reconstruction degrades gracefully as the SNR decreases.

\section{Experimental result}

\begin{figure}[htb]
\centering
\includegraphics[width=0.65\textwidth]{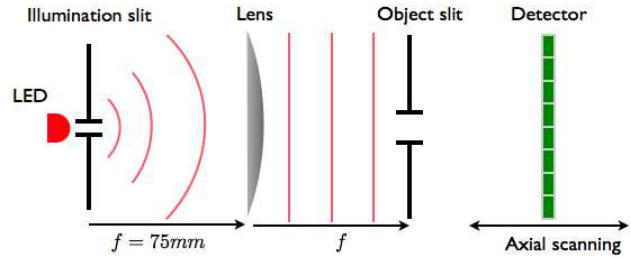}
\caption{Experimental arrangement}
\label{fig:setup}
\end{figure}

\begin{figure}[h]
\centering
\includegraphics[width=0.7\textwidth]{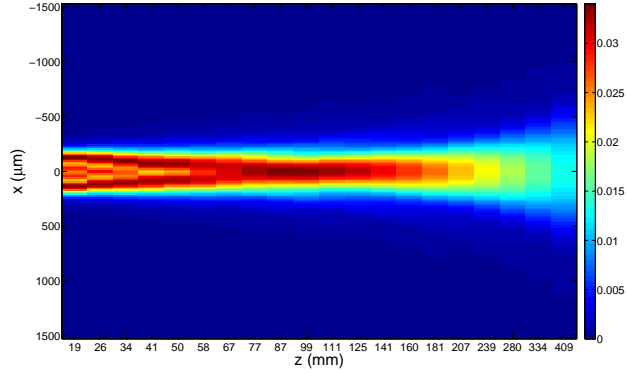}
\caption{Intensity measurements at several unequally spaced propagation distances.}
\label{fig:data}
\end{figure}

\begin{figure}[h]
\centering
\subfloat[]{\includegraphics[width=0.5\textwidth]{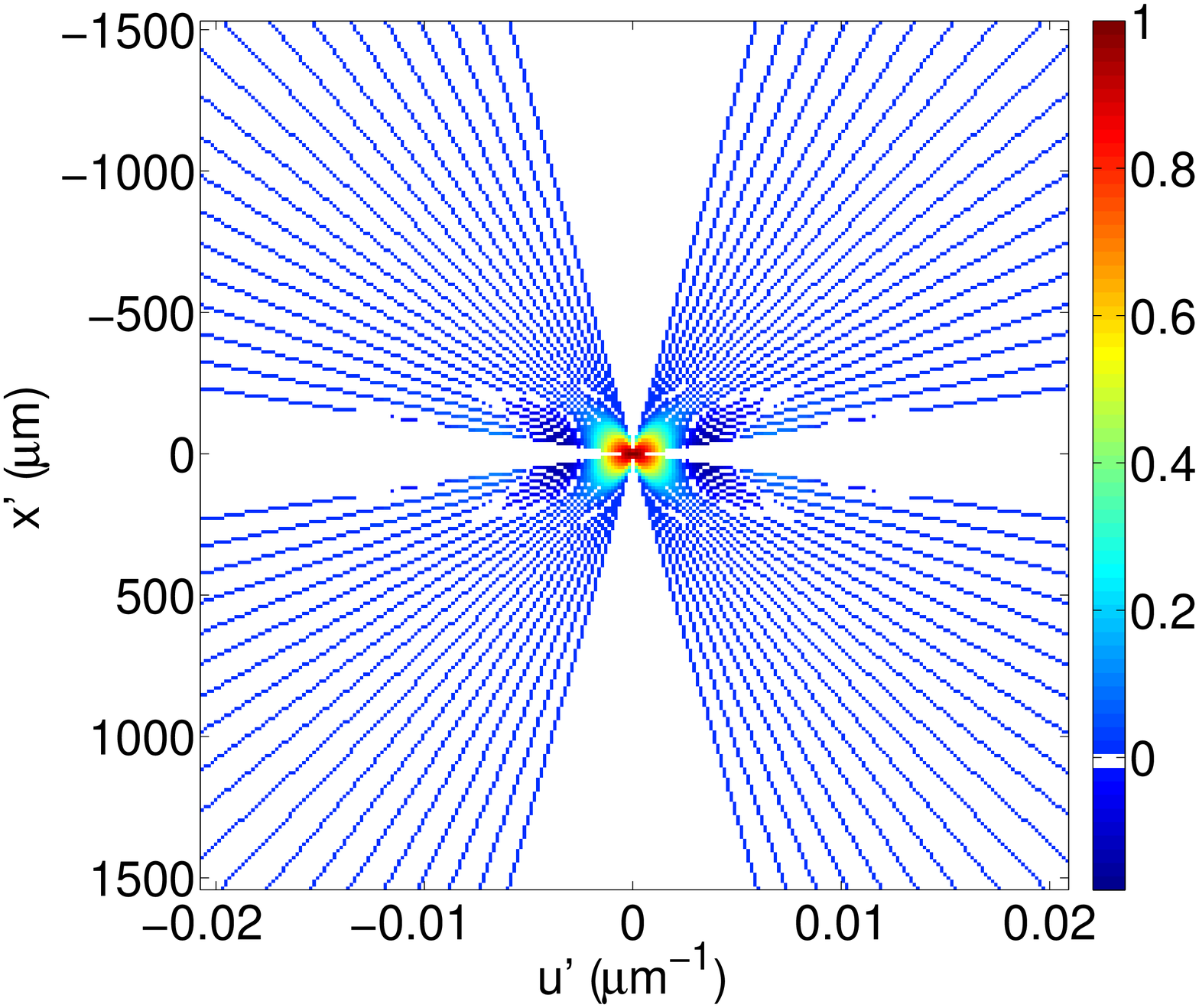}}
\subfloat[]{\includegraphics[width=0.5\textwidth]{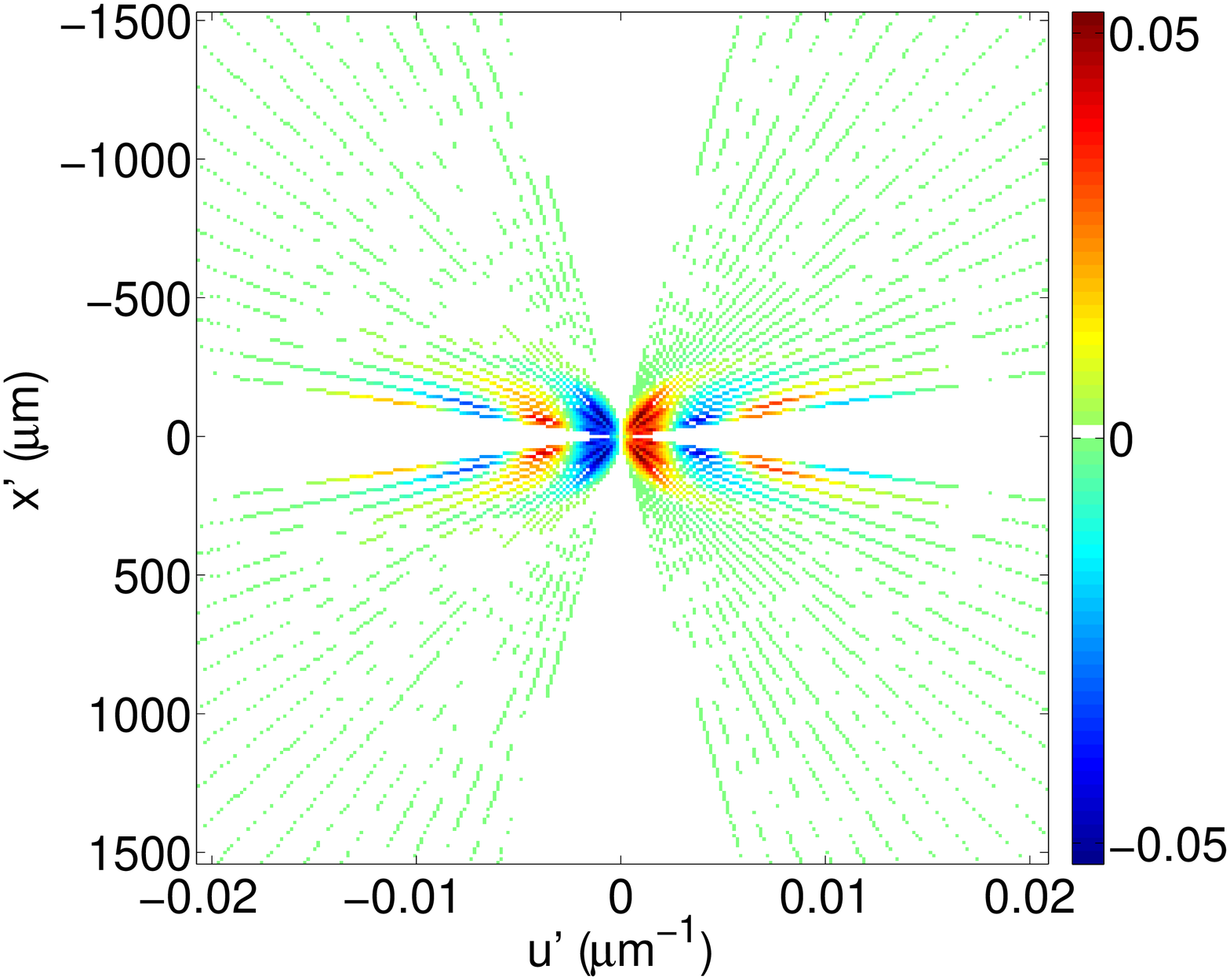}} 
\caption{(a) Real and (b) imaginary parts of the radial slices in Ambiguity space from Fourier transforming the vectors of intensities measured at corresponding propagation distances.}
\label{fig:AF}
\end{figure}

The experimental arrangement is illustrated in Fig.~\ref{fig:setup}. The illumination is generated by an LED with 620nm central wavelength and 20nm bandwidth. To generate partially coherent illumination, a single slit of width $355.6\mu\text{m}$ $(0.014'')$ is placed immediately after the LED and one focal length (75 mm) to the left of a cylindrical lens. One focal length to the right of the lens, we place the second single slit of width $457.2\mu\text{m}$ $(0.018'')$, which is used as a one--dimensional (1D) object.

The goal is to retrieve the mutual intensity immediately to the right of the object from a sequence of intensity measurements at varying $z$--distances downstream from the object, as described in the theory. We measured the intensities at 20 $z$--distances, ranging from 18.2mm to 467.2mm, to the right of the object. The data are given in Fig. \ref{fig:data}. Each 1D intensity measurement consists of 512 samples, captured by a CMOS sensor with $12\mu\text{m}$ pixel size. The dimension of the unknown mutual intensity matrix to be recovered is $512\times512$. Since only intensities at positive $z$, {\em i.e.} downstream from the object, are accessible, we can only fill up the top right and bottom left quadrants of Ambiguity space. The other two quadrants are filled symmetrically, {\em i.e.} assuming that if the field propagating to the right of the object were phase conjugated with respect to the axial variable $z$, it would yield the correct field to the left of the object, {\em i.e.} negative $z$~\cite{Tran:05, 1997Natur.386..150K}. Under this assumption, a total of 40 radial slices are sampled in Ambiguity space, as shown in Fig.~\ref{fig:AF}. The apex angle of the missing cone around the $u'$--axis is approximately 17.4 degrees, and the one around the $x'$--axis is approximately 28.6 degrees. The number of measurements is only $7.8\%$ of the total number of  entries in the unknown mutual intensity matrix.

\begin{figure}[htb]
\centering
\subfloat[]{\includegraphics[width=0.5\textwidth]{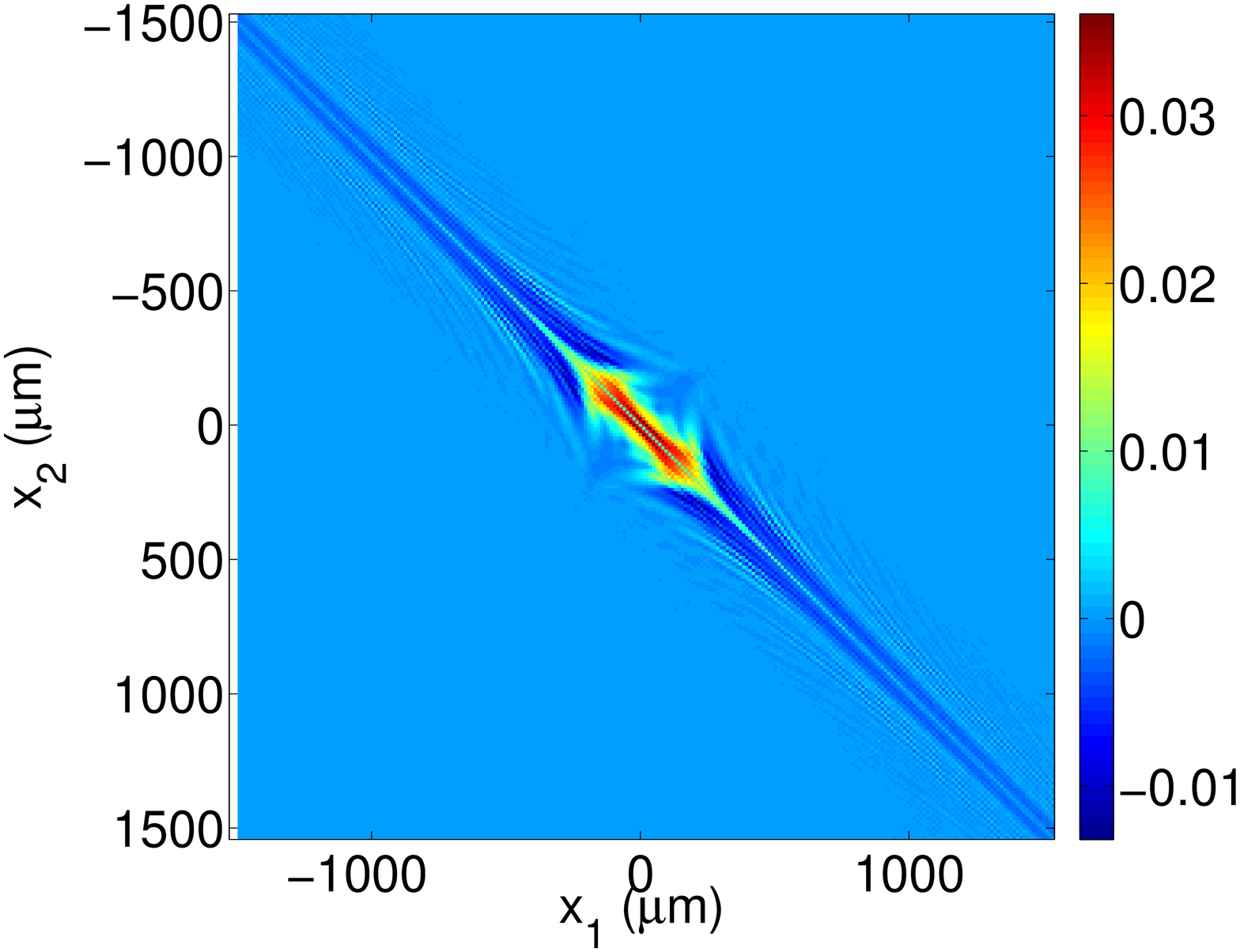}}
\subfloat[]{\includegraphics[width=0.5\textwidth]{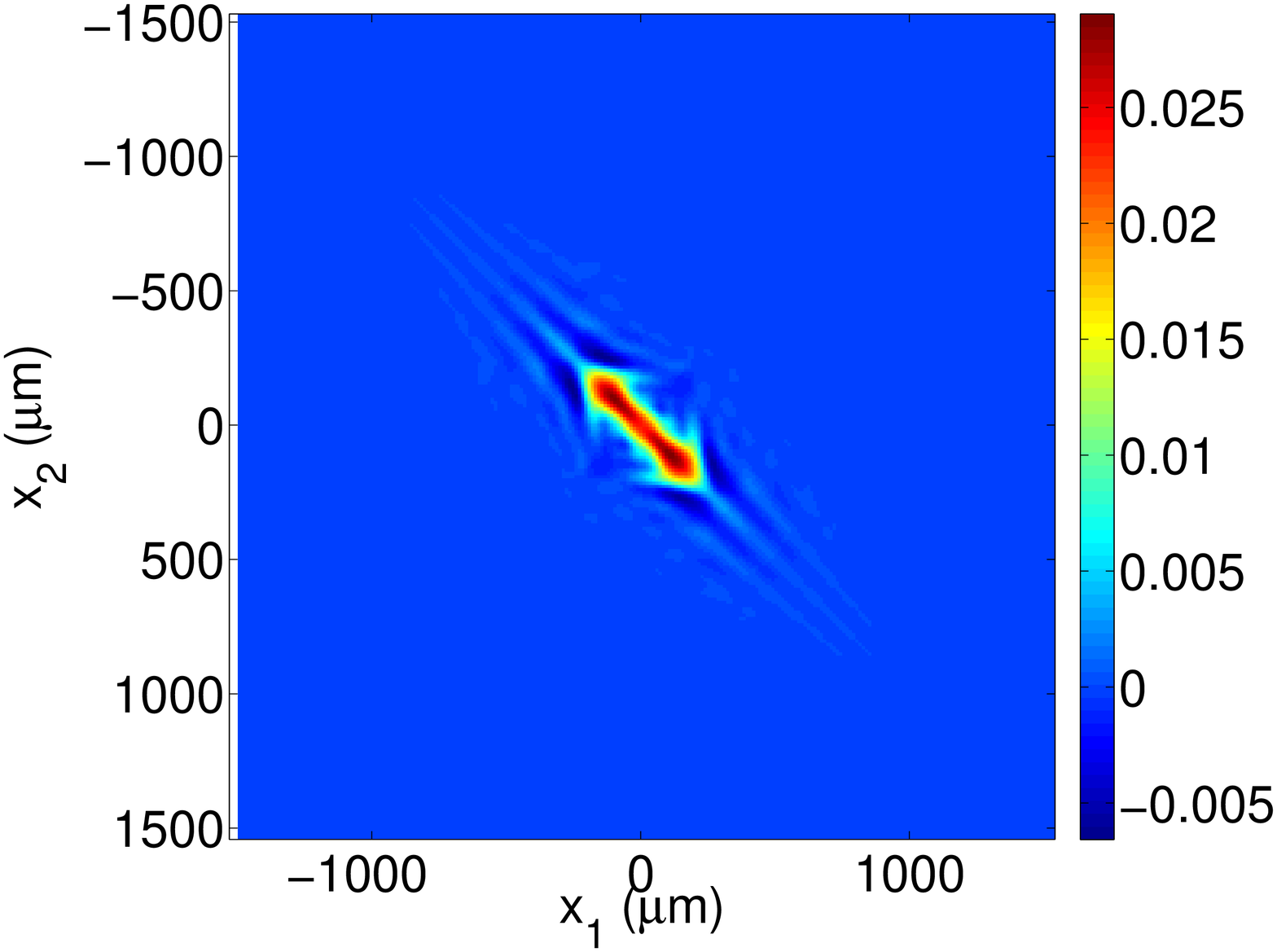}}
\caption{Real part of the reconstructed mutual intensity from (a) FBP; (b) LRMR method.}
\label{fig:MI_reconstruct}
\end{figure}

\begin{figure}[htb]
\centering
\subfloat[]{\includegraphics[width=0.5\textwidth]{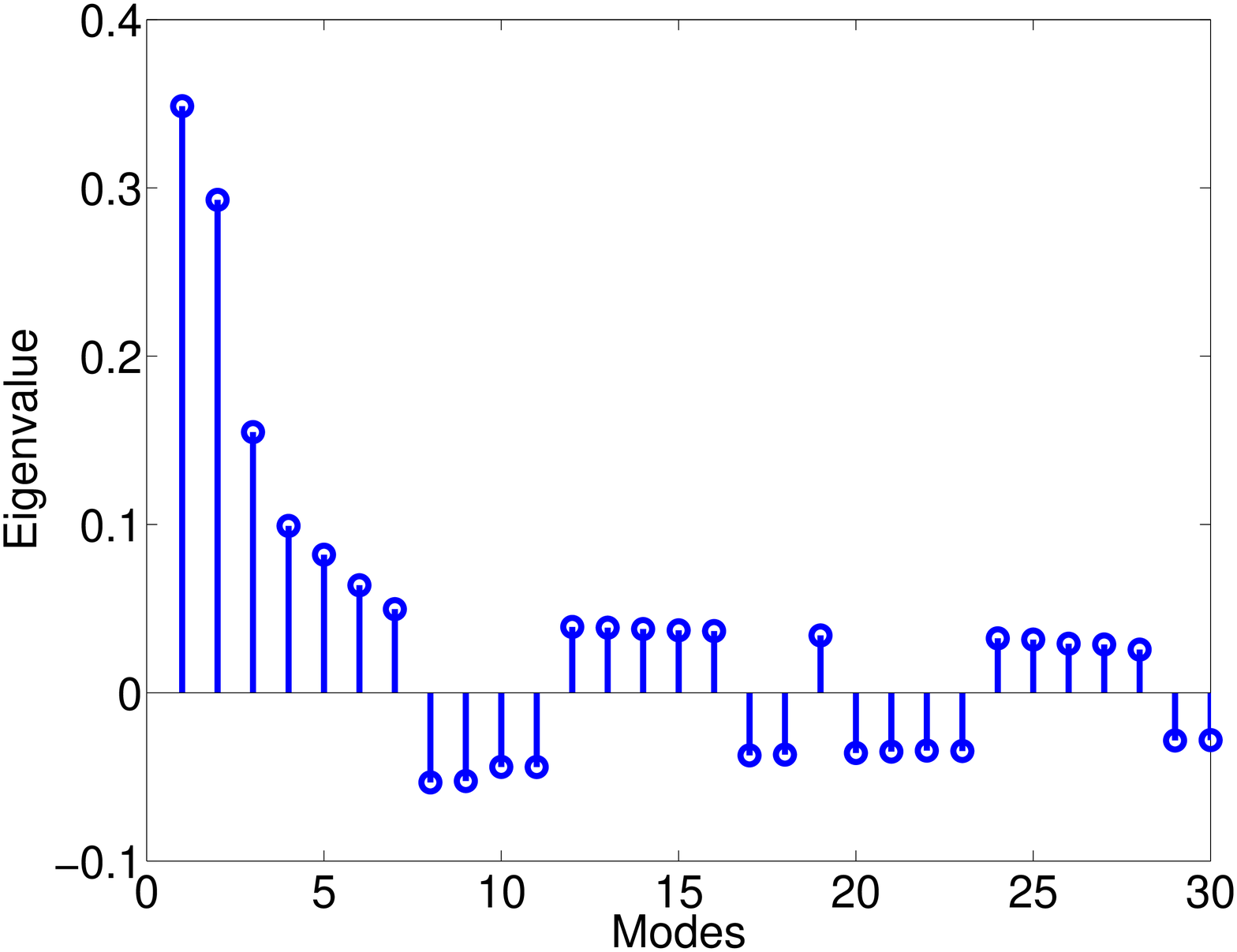}}
\subfloat[]{\includegraphics[width=0.5\textwidth]{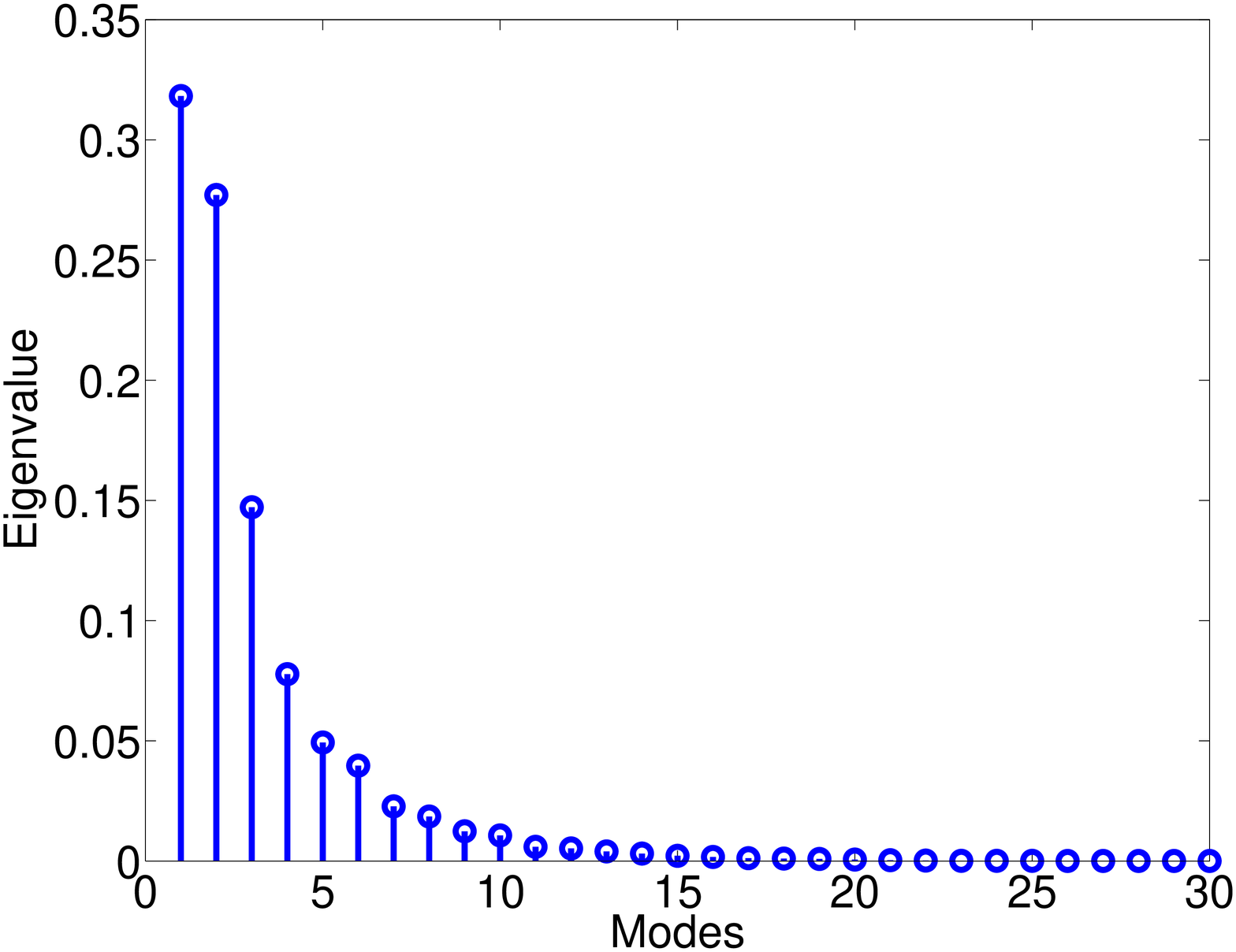}}
\caption{Eigenvalues estimated by (a) FBP, and (b) LRMR method.}
\label{fig:CMD}
\end{figure}

The reconstructions from the FBP and LRMR methods are compared in Fig.~\ref{fig:MI_reconstruct}(a) and \ref{fig:MI_reconstruct}(b), respectively. The FBP reconstruction suffers from the same artifacts detailed in the numerical simulations section. All these artifacts are greatly suppressed or completely removed in the LRMR reconstruction. In the real part of the reconstruction, the width of the square at the center is approximately $456\mu\text{m}$ (38 pixels), which agrees with the actual width of the slit. The imaginary part is orders of magnitude smaller than the real part. 

FBP estimated eigenvalues contain several negative values, and are shown in Fig.~\ref{fig:CMD}(a). This does not satisfy the positive energy constraint. LRMR estimated eigenvalues are compared in Fig.~\ref{fig:CMD}(b), and all eigenvalues are positive.

\begin{figure}[!]
\centering
\subfloat[]{\includegraphics[width=0.45\textwidth]{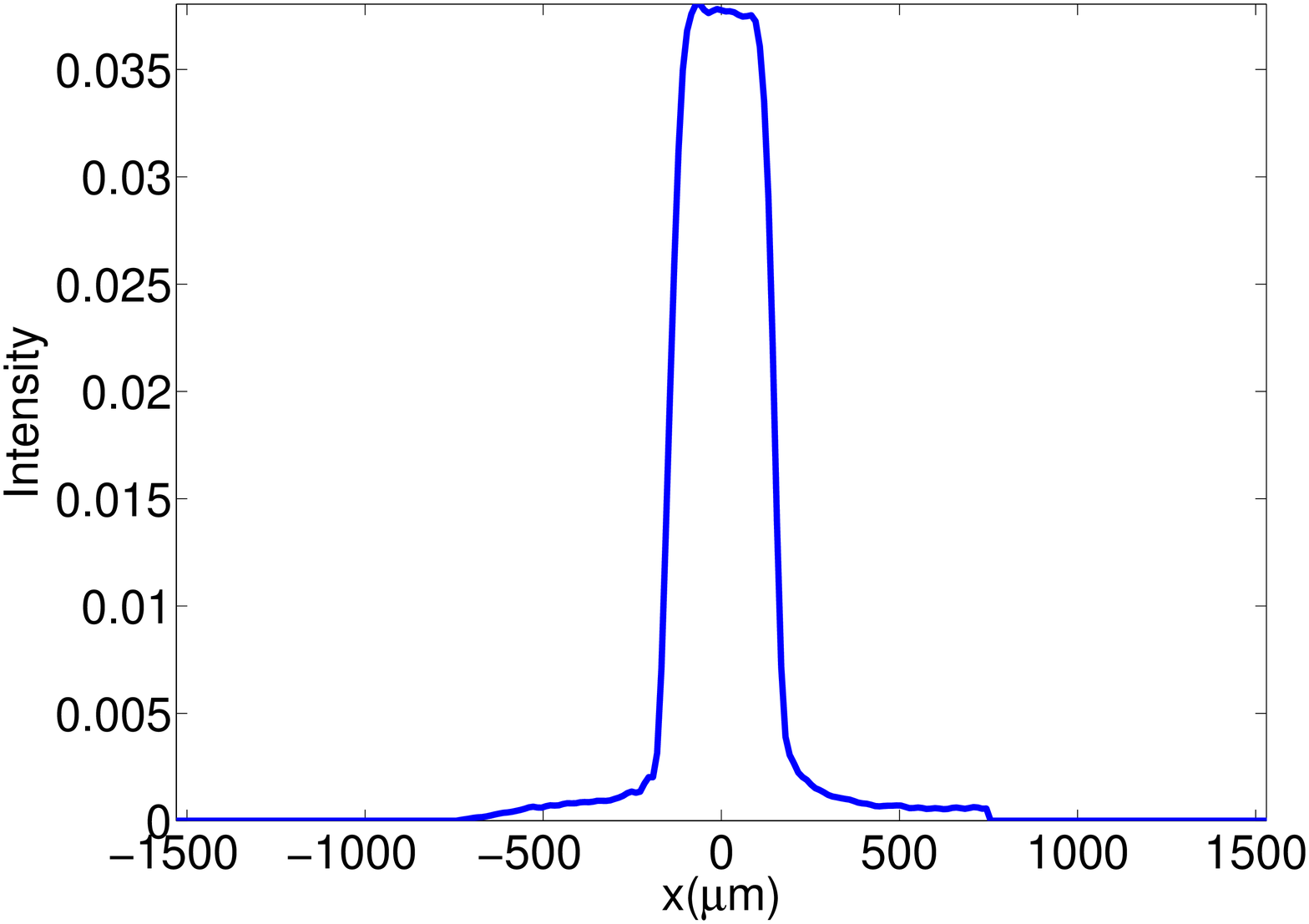}} 
\subfloat[]{\includegraphics[width=0.45\textwidth]{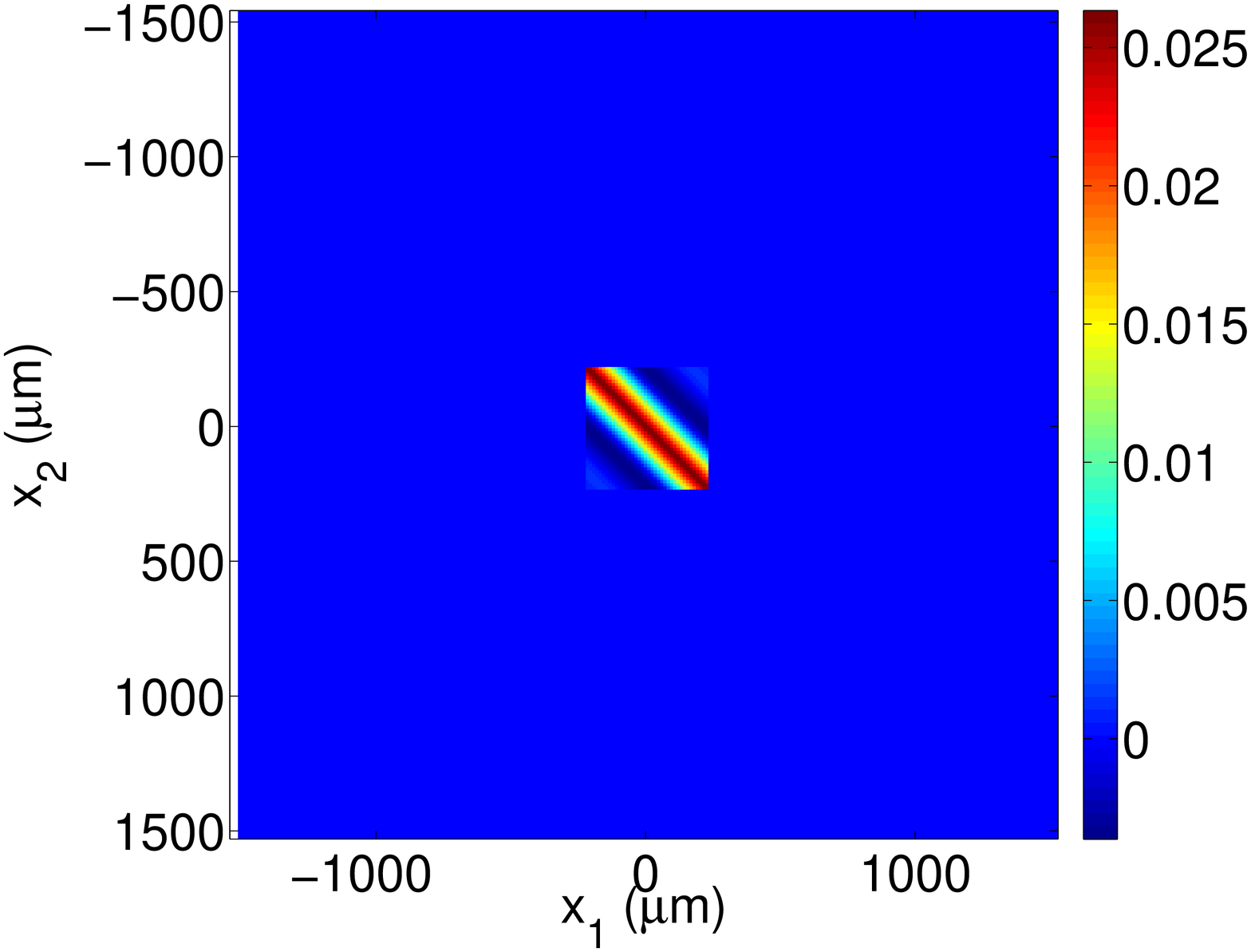}}

\subfloat[]{\includegraphics[width=0.45\textwidth]{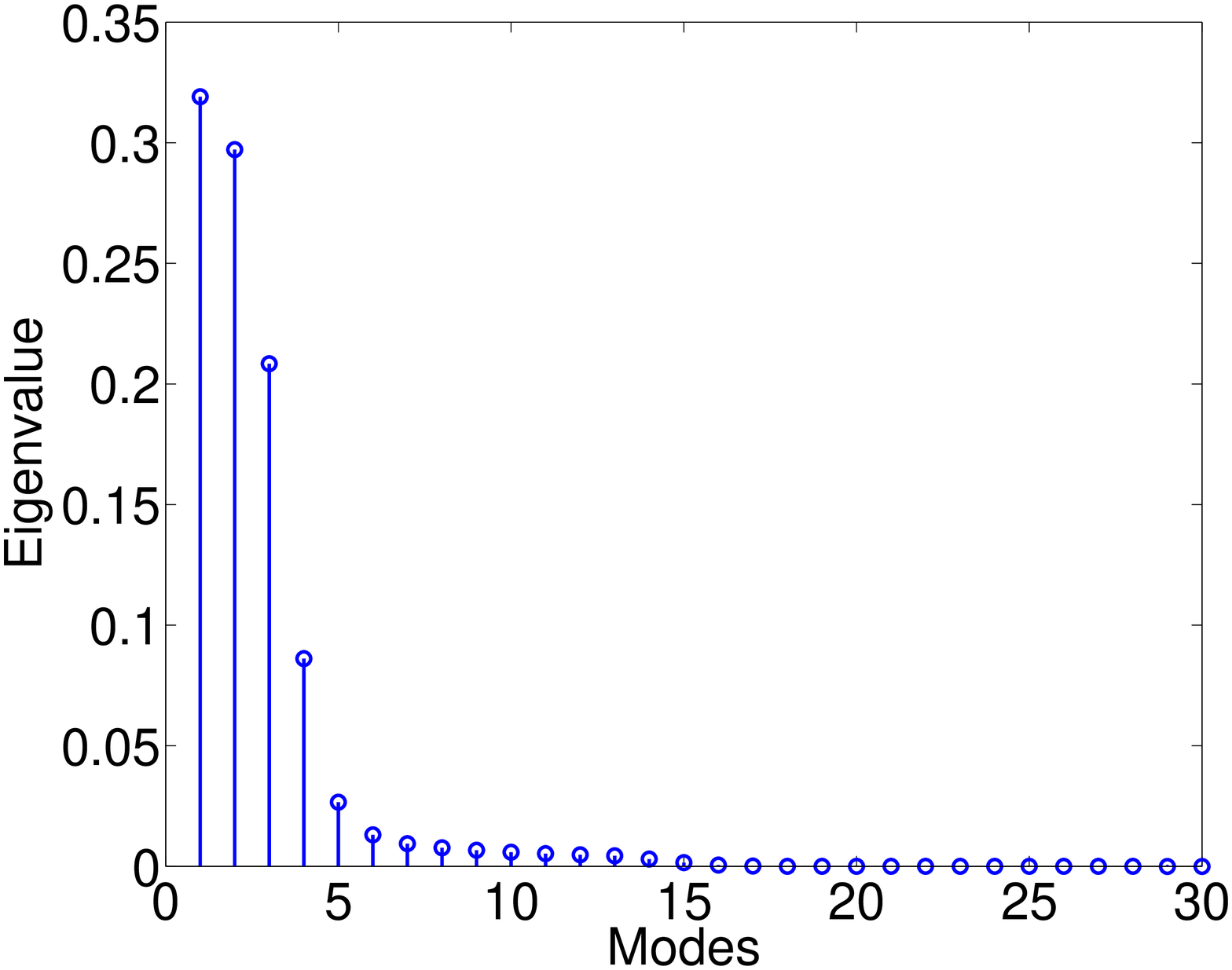}}
\subfloat[]{\includegraphics[width=0.45\textwidth]{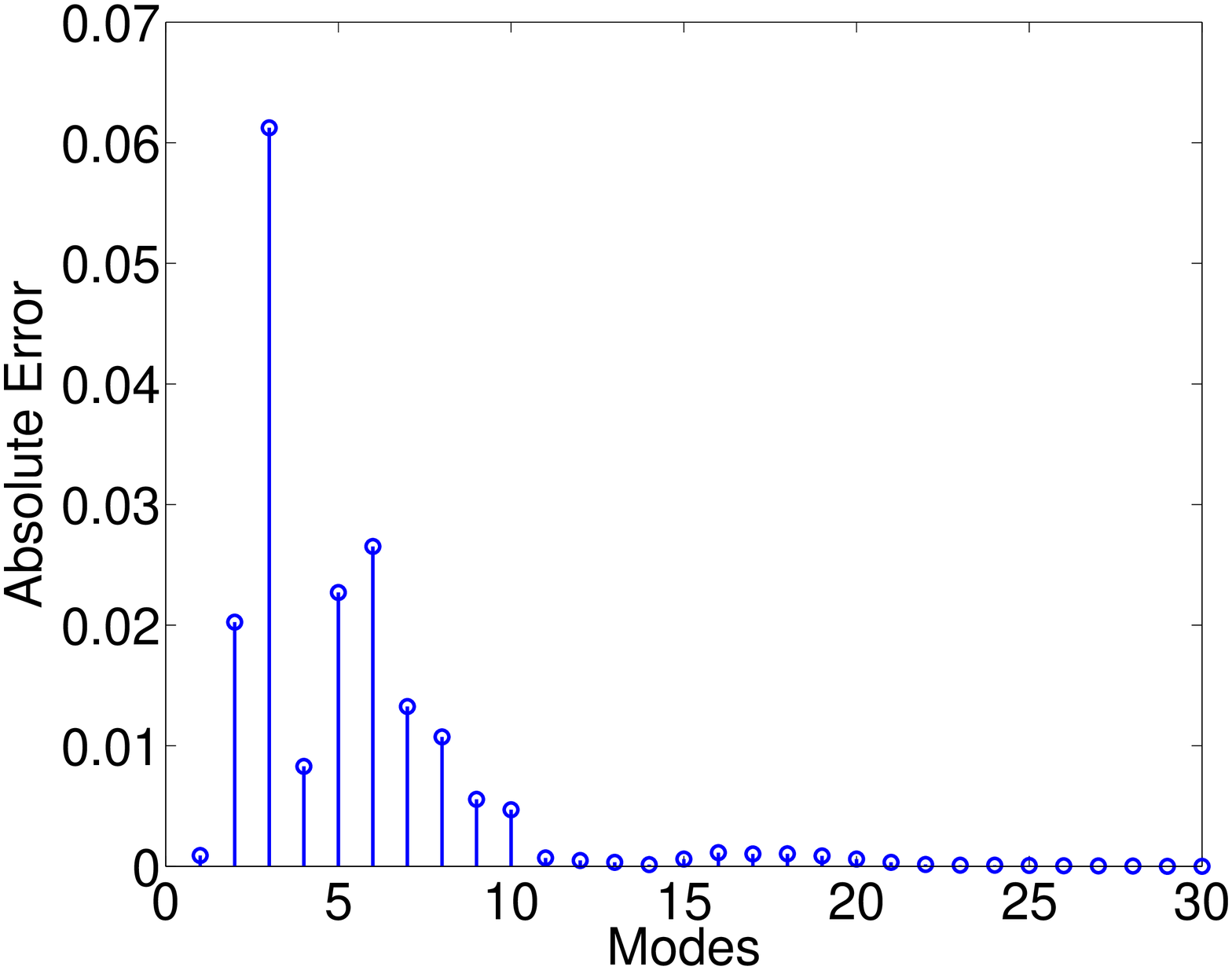}}
\caption{(a) Intensity measured immediately to the right of the illumination slit; (b) real part of van Cittert--Zernike theorem estimated mutual intensity immediately to the right of the object slit; (c) eigenvalues of the mutual intensity in (b); (d) absolute error between the eigenvalues in Fig.~\ref{fig:CMD}(b) and \ref{fig:Validate}(c) versus mode index.}
\label{fig:Validate}
\end{figure}

\begin{figure}[h]
\centering
\subfloat[]{\includegraphics[width=0.8\textwidth]{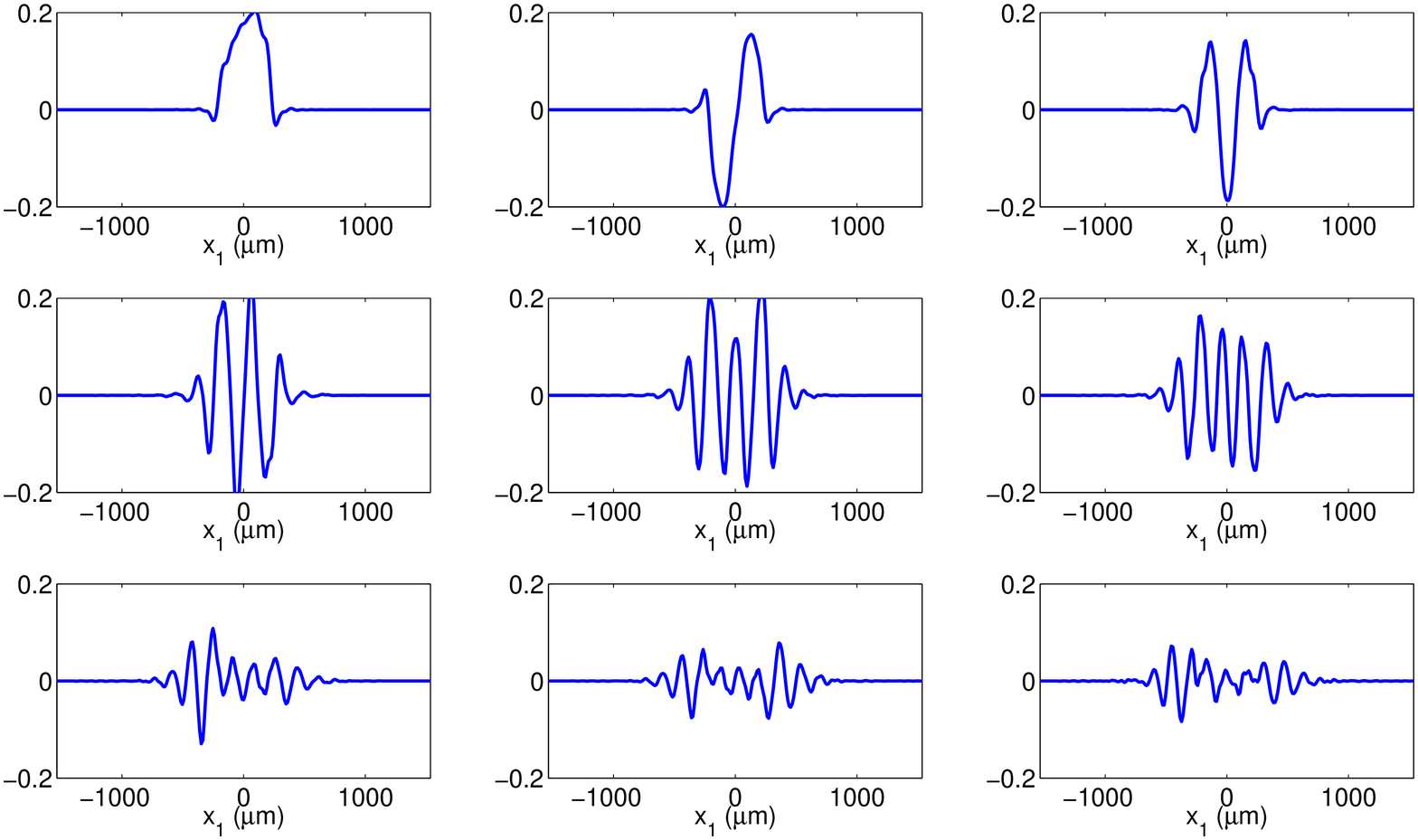}} 

\subfloat[]{\includegraphics[width=0.8\textwidth]{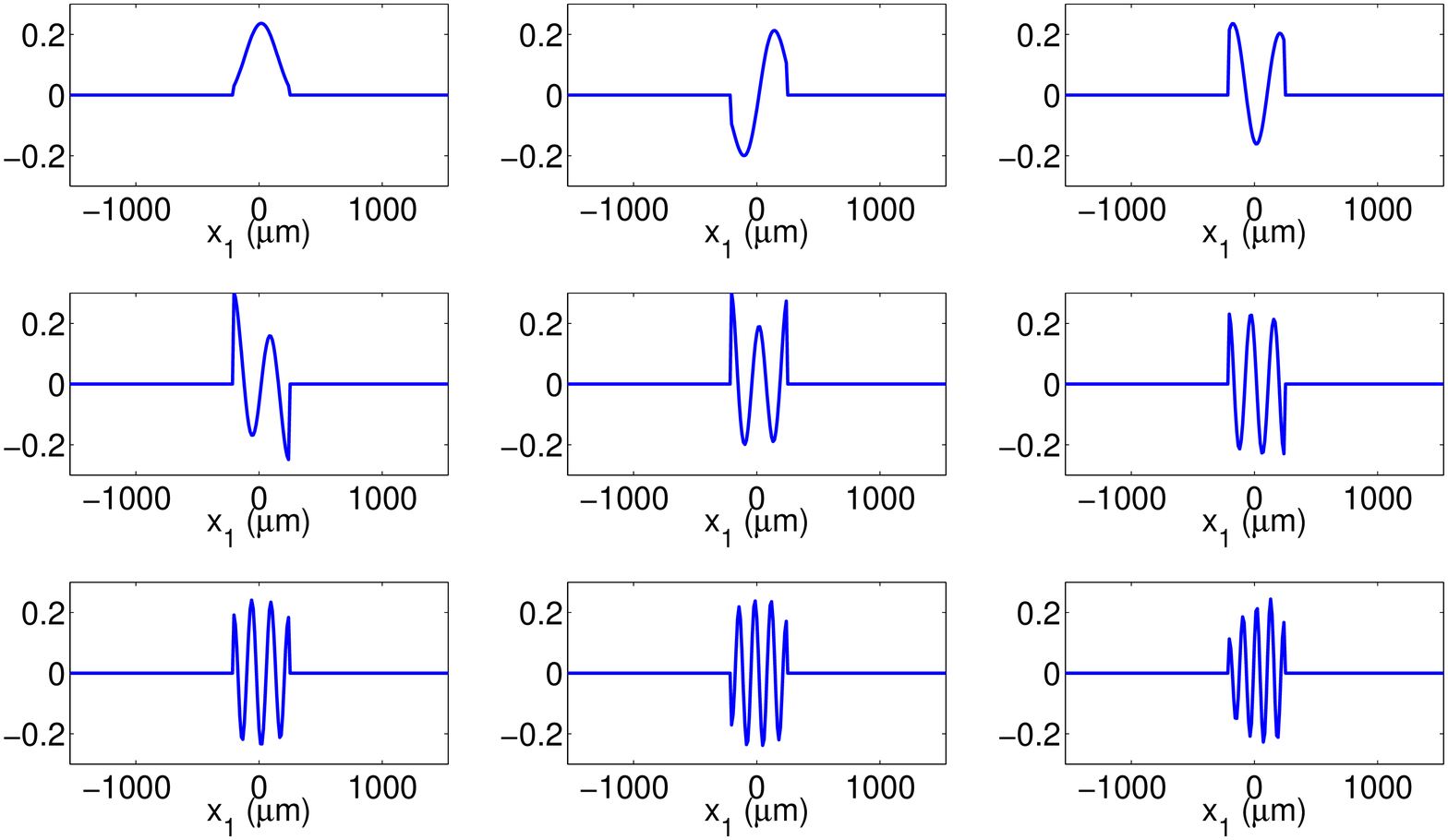}}
\caption{(a) LRMR estimated coherent modes of the mutual intensities in Fig.~\ref{fig:MI_reconstruct}(b), and (b) coherent modes of the mutual intensities in Fig.~\ref{fig:Validate}(b), calculated via use of the van Cittert--Zernike theorem, and assumption of incoherent illumination.}
\label{fig:eigenvectors}
\end{figure}

We further validated our compressive estimates by measuring the field intensity immediately to the right of the illumination slit [Fig.~\ref{fig:Validate}(a)]. Assuming that the illumination is spatially incoherent (a good assumption in the LED case), the mutual intensity of the field immediately to the left of the object is the Fourier transform of the measured intensity, according to the van Cittert--Zernike theorem~\cite{mandel1995optical, 2000stop.book.....G}. This calculated mutual intensity, based on the measurement of Fig.~\ref{fig:Validate}(a) and screened by the object slit, is shown in Fig.~\ref{fig:Validate}(b). The eigenvalues computed by coherent mode decomposition are shown in Fig.~\ref{fig:Validate}(c) and are in good agreement with the LRMR estimates, as compared in Fig.~\ref{fig:Validate}(d). It is seen that $99\%$ of the energy is contained in the first 13 modes, which confirms our low--rank assumption. The FBP reconstruction may also be compared quantitatively to the compressive reconstruction in terms of the global degree of coherence parameters, which were experimentally found as 0.12 and 0.46, respectively; whereas the estimate yielded by the van Cittert--Zernike theorem is 0.49. The first nine eigenvectors of each individual mode are shown and compared in Fig.~\ref{fig:eigenvectors}. Small errors in the compressive estimate are because the missing cone is still not perfectly compensated by the compressive approach, and because of other experimental imperfections. 

In this classical experiment, we have the benefit that direct observation of the one--dimensional object is available; thus, we were able to carry out quantitative analysis of the accuracy of the compressive estimate. In the quantum analogue of measuring a complete quantum state, direct observation would have of course not been possible, but the accuracy attained through the compressive estimate should be comparable, provided the low entropy assumption holds~\cite{PhysRevLett.105.150401}.

\section{Discussion}

In conclusion, we experimentally demonstrated compressive reconstruction of the mutual intensity function of a classical partially coherent source using phase space tomography. By exploiting the physically justifiable assumption of a quasi--pure source, both measurement and post--processing dimensionality are greatly reduced. We used the van Cittert--Zernike theorem to estimate the true mutual intensity function as a way to cross--validate the compressive reconstruction, and found indeed good agreement.

Here we followed a much simplified version of the approach described in~\cite{Candes:2011fk} which showed that the complex operators describing the measurements should be uniformly distributed in the $n$--dimensional unit sphere, whereas we simply utilized free space propagation. The phase masks described in~\cite{Candes:2011fk} to implement optimal sampling are outside the scope of the present work.

\section*{Acknowledgment}
The authors thank Baile Zhang, Jonathan Petruccelli, and Yi Liu for helpful discussions. Financial support was provided by Singapore's National Research Foundation through the Center for Environmental Sensing and Modeling (CENSAM) and BioSystems and bioMechanics (BioSyM) independent research groups of the Singapore-MIT Alliance for Research and Technology (SMART) Centre, by the US National Institutes of Health and by the Chevron--MIT University Partnership Program. 


\begin{thebibliography}{10}
\newcommand{\enquote}[1]{``#1''}

\bibitem{mandel1995optical}
L.~Mandel and E.~Wolf, \emph{Optical Coherence and Quantum Optics} (Cambridge
  University Press, 1995).

\bibitem{densitymatrix}
K.~Blum, \emph{Density Matrix Theory and Applications} (Plenum Press, 1981).

\bibitem{Itoh:86}
K.~Itoh and Y.~Ohtsuka, \enquote{Fourier-transform spectral imaging: retrieval
  of source information from three-dimensional spatial coherence,} J. Opt. Soc.
  Am. A \textbf{3}, 94--100 (1986).

\bibitem{Marks:99}
D.~L. Marks, R.~A. Stack, and D.~J. Brady, \enquote{{Three-dimensional
  coherence imaging in the Fresnel domain},} Appl. Opt. \textbf{38}, 1332--1342
  (1999).

\bibitem{2000stop.book.....G}
J.~W. {Goodman}, \emph{{Statistical Optics}} (Wiley-Interscience, 2000).

\bibitem{PhysRevLett.68.2261}
K.~A. Nugent, \enquote{Wave field determination using three-dimensional
  intensity information,} Phys. Rev. Lett. \textbf{68}, 2261--2264 (1992).

\bibitem{PhysRevLett.72.1137}
M.~G. Raymer, M.~Beck, and D.~McAlister, \enquote{Complex wave-field
  reconstruction using phase-space tomography,} Phys. Rev. Lett. \textbf{72},
  1137--1140 (1994).

\bibitem{Tran:05}
C.~Q. Tran, A.~G. Peele, A.~Roberts, K.~A. Nugent, D.~Paterson, and I.~McNulty,
  \enquote{X-ray imaging: a generalized approach using phase-space tomography,}
  J. Opt. Soc. Am. A \textbf{22}, 1691--1700 (2005).

\bibitem{Beck:93}
M.~Beck, M.~G. Raymer, I.~A. Walmsley, and V.~Wong, \enquote{Chronocyclic
  tomography for measuring the amplitude and phase structure of optical
  pulses,} Opt. Lett. \textbf{18}, 2041--2043 (1993).

\bibitem{1989PhRvA..40.2847V}
K.~{Vogel} and H.~{Risken}, \enquote{{Determination of quasiprobability
  distributions in terms of probability distributions for the rotated
  quadrature phase},} \pra \textbf{40}, 2847--2849 (1989).

\bibitem{PhysRevLett.70.1244}
D.~T. Smithey, M.~Beck, M.~G. Raymer, and A.~Faridani, \enquote{{Measurement of
  the Wigner distribution and the density matrix of a light mode using optical
  homodyne tomography: application to squeezed states and the vacuum},} Phys.
  Rev. Lett. \textbf{70}, 1244--1247 (1993).

\bibitem{PhysRevLett.74.4101}
U.~Leonhardt, \enquote{{Quantum--state tomography and discrete Wigner
  function},} Phys. Rev. Lett. \textbf{74}, 4101--4105 (1995).

\bibitem{1997Natur.386..150K}
C.~{Kurtsiefer}, T.~{Pfau}, and J.~{Mlynek}, \enquote{{Measurement of the
  Wigner function of an ensemble of Helium atoms},} \nat \textbf{386}, 150--153
  (1997).

\bibitem{candes2006robust}
E.~Cand{\`e}s, J.~Romberg, and T.~Tao, \enquote{Robust uncertainty principles:
  exact signal reconstruction from highly incomplete frequency information,}
  IEEE Trans. Inform. Theory \textbf{52}, 489--509 (2006).

\bibitem{Candes2006}
E.~{Cand{\`e}s}, J.~{Romberg}, and T.~{Tao}, \enquote{{Stable signal recovery
  from incomplete and inaccurate measurements},} Comm. Pure Appl. Math.
  \textbf{59}, 1207--1223 (2006).

\bibitem{Donoho:2006sf}
D.~L. Donoho, \enquote{Compressed sensing,} IEEE Trans. Inform. Theory
  \textbf{52}, 1289--1306 (2006).

\bibitem{Candes:fk}
E.~J. Cand{\`e}s and B.~Recht, \enquote{Exact matrix completion via convex
  optimization,} Found. Comput. Math. \textbf{9}, 717--772 (2009).

\bibitem{Candes:2010:PCR:1823677.1823678}
E.~J. {Cand{\`e}s} and T.~Tao, \enquote{The power of convex relaxation:
  near-optimal matrix completion,} IEEE Trans. Inform. Theory \textbf{56},
  2053--2080 (2010).

\bibitem{PhysRevLett.105.150401}
D.~Gross, Y.-K. Liu, S.~T. Flammia, S.~Becker, and J.~Eisert, \enquote{Quantum
  state tomography via compressed sensing,} Phys. Rev. Lett. \textbf{105},
  150401 (2010).

\bibitem{5714248}
D.~Gross, \enquote{Recovering low-rank matrices from few coefficients in any
  basis,} IEEE Trans. Inf. Theory \textbf{57}, 1548 --1566 (2011).

\bibitem{Candes:2011fk}
E.~J. Cand{\`e}s, T.~Strohmer, and V.~Voroninski, \enquote{Phaselift: exact and
  stable signal recovery from magnitude measurements via convex programming,}
  ArXiv: 1109.4499v1  (2011).

\bibitem{2011arXiv1109.0573C}
E.~J. {Cand{\`e}s}, Y.~{Eldar}, T.~{Strohmer}, and V.~{Voroninski},
  \enquote{{Phase retrieval via matrix completion},} ArXiv: 1109.0573  (2011).

\bibitem{Shechtman:11}
Y.~Shechtman, Y.~C. Eldar, A.~Szameit, and M.~Segev, \enquote{Sparsity based
  sub-wavelength imaging with partially incoherent light via quadratic
  compressed sensing,} Opt. Express \textbf{19}, 14807--14822 (2011).

\bibitem{Wolf:82}
E.~Wolf, \enquote{{New theory of partial coherence in the space-frequency
  domain. Part I: spectra and cross spectra of steady-state sources},} J. Opt.
  Soc. Am. \textbf{72}, 343--351 (1982).

\bibitem{Pelliccia:11}
D.~Pelliccia, A.~Y. Nikulin, H.~O. Moser, and K.~A. Nugent,
  \enquote{Experimental characterization of the coherence properties of hard
  x-ray sources,} Opt. Express \textbf{19}, 8073--8078 (2011).

\bibitem{Brenner1983323}
K.-H. Brenner, A.~Lohmann, and J.~Ojeda-Casta{\~n}eda, \enquote{{The ambiguity
  function as a polar display of the OTF},} Opt. Commun. \textbf{44}, 323 --
  326 (1983).

\bibitem{1984AcOpt..31..213B}
K.-H. Brenner and J.~{Ojeda-Casta{\~n}eda}, \enquote{{Ambiguity function and
  Wigner distribution function applied to partially coherent imagery},} Opt.
  Acta. \textbf{31}, 213--223 (1984).

\bibitem{Tu:1997fk}
J.~Tu, \enquote{Wave field determination using tomography of the ambiguity
  function,} Phys. Rev. E \textbf{55}, 1946--1949 (1997).

\bibitem{2009arXiv0903.3131C}
E.~J. {Cand{\`e}s} and Y.~{Plan}, \enquote{{Matrix completion with noise},}
  ArXiv: 0903.3131  (2009).

\bibitem{2008arXiv0810.3286C}
J.-F. {Cai}, E.~J. {Cand{\`e}s}, and Z.~{Shen}, \enquote{{A singular value
  thresholding algorithm for matrix completion},} ArXiv: 0810.3286  (2008).

\bibitem{Starikov:82}
A.~Starikov and E.~Wolf, \enquote{{Coherent-mode representation of Gaussian
  Schell-model sources and of their radiation fields},} J. Opt. Soc. Am.
  \textbf{72}, 923--928 (1982).

\bibitem{Kak:1988fk}
A.~C. Kak and M.~Slaney, \emph{Principle of Computerized Tomographic Imaging}
  (Society for Industrial and Applied Mathematics, 2001).

\bibitem{1978OptCo..25...26B}
M.~J. {Bastiaans}, \enquote{{The Wigner distribution function applied to
  optical signals and systems},} Opt. Commun. \textbf{25}, 26--30
  (1978).

\bibitem{Bastiaans:86}
M.~J. Bastiaans, \enquote{{Application of the Wigner distribution function to
  partially coherent light},} J. Opt. Soc. Am. A \textbf{3}, 1227--1238 (1986).

\bibitem{Starikov:82mode}
A.~Starikov, \enquote{Effective number of degrees of freedom of partially
  coherent sources,} J. Opt. Soc. Am. \textbf{72}, 1538--1544 (1982).

\bibitem{Bastiaans:84}
M.~J. Bastiaans, \enquote{New class of uncertainty relations for partially
  coherent light,} J. Opt. Soc. Am. A \textbf{1}, 711--715 (1984).

\end{thebibliography}
\end{document}